\documentclass[12pt, draftclsnofoot, onecolumn]{IEEEtran}

\renewcommand{\baselinestretch}{1.40}

\usepackage{cite}
\usepackage{algorithmic}
\usepackage{array}
\usepackage{bbm}
\usepackage{amsfonts}
\usepackage{graphicx}
\usepackage{xcolor}
\usepackage{mathtools}
\usepackage{subfigure,epsfig}
\usepackage{amsmath, amsthm, amssymb}
\usepackage[outdir=./]{epstopdf}
\usepackage{dsfont}
\usepackage{mathrsfs}
\newtheorem{theorem}{Theorem}
\newtheorem{Corollary}{Corollary}
\newtheorem{Definition}{Definition}

\def\SIR{\text{SIR}}
\def\dbar#1{\bar{\bar{#1}}}
\def\figref#1{Fig.\,\ref{#1}}%

\newlength{\figwidth}
\newcommand*{\Scale}[2][4]{\scalebox{#1}{$#2$}}%

\begin{document}
\title{Meta Distribution of the SIR in Large-Scale Uplink and Downlink NOMA Networks
}
\author{Mohammad Salehi, Hina Tabassum, and Ekram Hossain\thanks{The authors are with the Department of Electrical and Computer Engineering at the University of Manitoba, Canada.
 (emails: salehim@myumanitoba.ca, \{Hina.Tabassum, Ekram.Hossain\}@umanitoba.ca). 
 }}

\maketitle



\IEEEpeerreviewmaketitle

\begin{abstract}

We develop an analytical framework to derive the meta distribution and moments of the conditional success probability  (CSP), which is defined as {success probability for a given realization of the transmitters},  in large-scale co-channel uplink and downlink non-orthogonal multiple access (NOMA) networks with one NOMA cluster per cell. The moments of CSP translate to various network performance metrics such as the  standard success or signal-to-interference ratio (SIR) coverage probability (which is the $1$-st moment), the mean local delay (which is the $-1$-st moment in a static network setting), and the meta distribution (which is the complementary cumulative distribution function of the conditional success probability and can be approximated by using the $1$-st and $2$-nd moments). For  uplink NOMA, to make the framework tractable, we propose two point process models for the spatial locations of the interferers by utilizing the base station (BS)/user pair correlation function. We validate the proposed models by comparing  the second moment measure of each model  with that of the  actual point process for the inter-cluster (or inter-cell) interferers obtained via simulations. 
For downlink NOMA, we derive closed-form solutions for the moments of the CSP, success (or coverage) probability, average local delay, and meta distribution for the users. As an application of the developed analytical framework, we use the closed-form expressions to optimize the power allocations for downlink NOMA  users in order to maximize the success probability of a given NOMA user with and without  latency constraints.  Closed-form optimal solutions for the transmit powers are obtained for two-user NOMA scenario. We note that maximizing the success probability with latency constraints can significantly impact the optimal power solutions for low  SIR thresholds and favour orthogonal multiple access (OMA).
\end{abstract}


\section{Introduction}

The  next generations of wireless networks  (such as 5G~\cite{Hossain2014} or beyond 5G [B5G]) are expected  to  support  billions of devices that are stimulated  mainly from the diverse Internet-of-Things (IoT) applications  (ranging from delay-tolerant machine-type communications (MTC) to delay sensitive mission-critical communications) in addition to the enhanced mobile broadband applications. As a result, acquiring ultra-reliable and low-latency  communication (URLLC) is among one of the constitutional challenges for emerging massive wireless networks~\cite{bennis2018ultra}. Recently,  NTT DOCOMO and  Huawei jointly conduct a successful field trial focused on the URLLC use-case with a macro base station on the 4.5 GHz frequency band (C-Band) using a new radio interface of similar features such as 3GPP 5G New Radio (NR) air-interface. 
Traditionally, reliability can be achieved with  efficient channel coding and  retransmission schemes, e.g., hybrid automatic repeat request (HARQ). However, at the same time, massive device connectivity with strict latency requirements need to be achieved in URLLC systems. This fact necessitates efficient user access mechanisms that can potentially serve multiple devices in a specific time-frequency resource block while reducing their respective transmission delays~\cite{mostafa2017connectivity}.  

Non-orthogonal multiple access (NOMA) has been recognized as a promising multi-user channel access technique that enables massive connectivity while reducing the  transmission delay of the devices~\cite{mostafa2017connectivity}. Contrary to traditional orthogonal multiple access (OMA), such as time division multiple access (TDMA), frequency division multiple access (FDMA), and code division multiple access (CDMA),  the key idea of NOMA is to serve multiple users in the same channel simultaneously. The concurrent transmissions in NOMA shorten the waiting time of the devices  while saving network resources. Of course, this can be achieved at the expense of additional interference and decoding complexity at the receivers. In particular, to mitigate the interference, NOMA exploits successive interference cancellation (SIC) at the receivers~\cite{saito2013,noma}.

\subsection{Background Work}
Recently, performance analysis of NOMA-based wireless networks  has attracted significant research interest. The existing studies contribute mainly toward understanding the {\bf average performance} of users  considering a single NOMA cell/cluster~\cite{Ding2014,dingearly,noma,zhang2016v1,uplink1,uplink2}. For instance, the performance of a single-cell downlink NOMA system with randomly located users was first studied in \cite{Ding2014}. In particular, the signal-to-interference-plus-noise ratio (SINR) outage probability and the ergodic capacity were derived for a user at rank $m$ in terms of distance. In \cite{dingearly}, the problem of user pairing was investigated  considering  fixed NOMA (F-NOMA) and cognitive radio inspired (CR-NOMA). In F-NOMA, any two users can make a NOMA pair  based on their channel gains.  On the other hand, in CR-NOMA, a weak channel user opportunistically gets paired with the strong channel user provided  that the interference caused by the strong user does not harm the weak channel user. A comparative performance analysis of uplink and downlink NOMA with selective two user pairing was conducted in \cite{noma}.  Closed-form solutions for ergodic sum-rate and outage probability of  a two-user NOMA cluster were presented in \cite{zhang2016v1} considering  a power back-off policy.  The power back-off policy was applied to distinguish users in a NOMA cluster with nearly similar signal strengths (given that traditional uplink power control is in effect). The problem of user scheduling, subcarrier allocation, and power control in uplink NOMA was investigated in~\cite{uplink1,uplink2} with perfect SIC at the BS. 


The aforementioned research studies ignore the  impact of inter-cell interference which can significantly limit the performance of NOMA in massive wireless networks. Very recently, some of the research works have considered the performance characterization of large-scale NOMA systems using tools  such as Poisson point process (PPP) and Poisson cluster process (PCP) from stochastic geometry.  The performance of uplink NOMA in terms of the rate coverage and average achievable rate was characterized first in \cite{Tabassum2017} using PCP considering both perfect and imperfect SIC. For downlink  NOMA, outage probability and average achievable rate of $m$-th rank user were derived in \cite{Zhang2016,Zhang2017} assuming that the  BS locations follow a homogeneous PPP. The  users are ranked based on their normalized channel gains defined as the channel gain including path loss and small-scale fading normalized by the inter-cell interference. The analytical expressions are derived assuming that {\em the normalized channel gains of users  located in a given NOMA cluster are independent and identically distributed (i.i.d.)}. However,  since the inter-cell interferences  received at the different users in the downlink are correlated,  the normalized channel gains are also correlated, and therefore,  the derived results are not precise.  
Another interesting work is \cite{Liu2017JSAC} where the performance of two-user downlink NOMA was investigated in a $K$-tier cellular network. The macro cell BSs use the massive multiple-input multiple-output (MIMO) technology and each small cell adopts user pairing to implement two-user NOMA transmission.
In \cite{Liu2017}, for $K$-tier heterogeneous networks (HetNets) with biased nearest BS association, performance of downlink NOMA was investigated in terms of the coverage probability and throughput for non-cooperative and cooperative schemes. 
{For Poisson cellular networks, \cite{Zhang2017} also studied the performance of uplink NOMA. To derive the analytical results, it was assumed that uplink interferers form a homogeneous PPP which is not correct.}

\subsection{Motivation and Contributions}

The current state-of-the-art mainly analyze the  {standard} transmission success probability and ergodic capacity of users in NOMA-enabled cellular networks.
Nevertheless, it is noteworthy that {the standard} transmission success probability is itself {the mean of a random variable referred to as {\em conditional success probability (CSP)}, which is the success probability of a user considering a given realization of  BSs~\cite{Haenggi2016}.} 
When the point process describing the receiver locations (referred to as receivers' point process) is ergodic, the {standard} success probability is the average of the {CSPs of all users}. Two networks can have the same {standard} (mean) success probability but distributions of the CSPs may be completely different. This is similar to the case where two different random variables have the same mean but different probability density functions (PDFs). Therefore, comparing two networks simply in terms of their average CSPs (or mean success probabilities) will not always be accurate since the CSP will not always be precisely characterized by its average value.

Along this line, \cite{Haenggi2016} characterized the {\em meta distribution} which is the {\em complementary cumulative distribution function (CCDF) of the CSP} by deriving the moments of the CSP. This pioneering work was followed by various  research studies for Poisson bipolar networks, device-to-device (D2D) networks, and millimeter-wave (mm-wave) D2D networks~\cite{Haenggi2016,Wang2017,Cui2017CoMP,Salehi2017,Deng2017}. The meta distribution provides a more precise characterization of a typical transmission link than the standard success probability and enables us to answer questions such as ``what fraction of users  ($y$) can be guaranteed with a coverage probability higher than a given target value of $x$?".  Cellular operators may be more interested in the performance level that $y$\% of  users achieve instead of the performance of a ``typical user".

To this end, our main contributions in this paper can be summarized as follows:
\begin{itemize}
\item We derive the moments of CSP for uplink and downlink NOMA in Poisson cellular networks. This allows us to study the traditional success/coverage probability (which is the $1$-st moment), the mean local delay (which is the $-1$-st moment), and the meta distribution (which is the CCDF of the success or SIR coverage probability and can be approximated using the $1$-st and $2$-nd moments).   Note that, {\em mean local delay}, which is defined as the mean number of transmission attempts until the first successful reception~\cite{Baccelli2010}, is a crucial performance metric for emerging URLLC systems.  

\item In uplink NOMA, the point process for the spatial locations of the interferers  is a key for the derivation of the meta distribution and moments of CSP. Since the actual point process is unknown, we propose two models for this point process based on the pair correlation between {interferers} and the typical BS (which is at the origin).  We  demonstrate the accuracy of the proposed point processes by comparing the  second moment measure\footnote{The matching of moment measures is different  from traditional moment matching of two random variables since it is the matching in two dimensions.} of each process with that of the original process obtained via simulations. 
We show that the proposed  point processes provide better approximations for low SIR threshold $\theta$, user locations closer to the BS, and dense BS deployments.

\item For downlink NOMA, we derive closed-form expressions for the moments of the CSP, success probability, average local delay, and the meta distribution. We approximate the meta distribution by a beta distribution and demonstrate the accuracy of the approximation.

\item As an application of the developed analytical framework, we use the closed-form results to optimize the power allocations for  downlink NOMA users with an objective to maximizing the success probability with and without latency constraints. The optimal solutions for the transmit powers are obtained in closed-form for the special case of two-user NOMA (i.e., two user per NOMA cluster). We note that maximizing the success probability with strict latency constraints can significantly impact the optimal power solutions for low  SIR thresholds and can favour OMA.

\end{itemize}

\subsection{Paper Organization and Notations}
The rest of the paper is structured as follows. Section~II briefly discusses the mathematical preliminaries related to the meta distribution, local delay, and their analytical evaluations. In Section~III, we describe the system model and assumptions for uplink and downlink NOMA. In Section~IV, for uplink NOMA, we propose two point processes to model the locations of the interferers and  derive the moments of the CSP and its meta distribution. In Section~V, for downlink NOMA, we derive closed-form solutions for the CSP and its meta distribution. Based on the closed-form solutions, in Section VI, we optimize the transmit powers for  the downlink NOMA users in order to maximize their success probabilities under latency constraints.  Section~VII discusses numerical and simulation results followed by the conclusion in Section~VIII. 

\section{Mathematical Preliminaries}
Consider a static cellular network where receivers are distributed according to a homogeneous Poisson Point Process (PPP) $\Phi_{\rm r}$. Because of the stationarity of the homogeneous PPP, we can condition on having a receiver at the origin {which is called the typical receiver}. We  denote the  distribution of transmitters with $\Phi$. For such a set-up, the concepts of CSP and the meta distribution along with their evaluation methods are defined in the following to provide a preliminary mathematical background to readers.

\begin{Definition}[Conditional Success Probability (CSP)~\cite{Haenggi2016}] Given the location of the transmitters and conditioned on the desired transmitter to be active, CSP {is} defined as follows:
\begin{IEEEeqnarray}{rCl}
	P_{\rm s}(\theta) &\triangleq& \mathbb{P}( \text{\rm SINR}>\theta \mid \Phi,\text{\rm tx}),
	\label{CSP}
\end{IEEEeqnarray}
where $\theta$ is the desired SINR and the $b$-th moment of $P_{\rm s} (\theta)$ {is} given by $M_b=\mathbb{E}_{\Phi}\left[ P_{\rm s}^b \right]$. 
\end{Definition}

\begin{Definition}[Meta Distribution of CSP]
Meta distribution is the CCDF of  $P_{\rm s} (\theta)$, {i.e.},
\begin{IEEEeqnarray}{rCl}
	\bar{F}_{P_{\rm s}}(x) &\triangleq& \mathbb{P}^{!0}( P_{\rm s}(\theta)>x ), \qquad x\in[0,1],
	\label{MD}
\end{IEEEeqnarray} 
in which $\mathbb{P}^{!0}$ is the reduced Palm measure given that the typical receiver is at the origin.
\end{Definition}
When $\Phi_{\rm r}$ is ergodic \cite{haenggi2012stochastic}, the meta distribution can be interpreted as the fraction of active users whose success probabilities are more than $x$ in each realization. In \cite{Haenggi2016}, an exact expression along with an approximation and simple bounds for the meta distribution were provided. A summary of these results is given below.
\begin{itemize}
\item {\em Exact meta distribution of CSP:} To derive the exact meta distribution, we first need to derive imaginary moments $M_{jt}=\mathbb{E}_{\Phi}\left[ P_{\rm s}^{jt} \right]$, where $j=\sqrt{-1}$ and $t\in\mathbb{R}^+$. Then using the Gil-Pelaez theorem \cite{Gil1951}, the exact meta distribution {is} given as follows:
\begin{IEEEeqnarray}{rCl}
	\bar{F}_{P_{\rm s}}(x) &=& \frac{1}{2} + \frac{1}{\pi} \int\limits_{0}^{\infty} \frac{ \Im\left(e^{-jt\log{x}} M_{jt}\right) }{t}{\rm d}t, \qquad x\in[0,1],
	\label{exactMD}
\end{IEEEeqnarray}
where $\Im(s)$ gives the imaginary part of $s$. 
\item {\em Approximate meta distribution of CSP:} A simple approximation of the meta distribution is provided by using the beta distribution. In this approach, we  need to derive the first moment $M_1$ and  the second moment $M_2$ of $P_s(\theta)$ and match them with the first and second moments of the beta distribution, i.e.,
\begin{IEEEeqnarray}{rCl}
	\bar{F}_{P_{\rm s}}(x) &\approx& 1-I_x\left( \frac{M_1 \beta}{1-M_1},\beta \right), \qquad x\in[0,1],
	\label{approxMD}
\end{IEEEeqnarray}  
where $\beta = \frac{ \left(M_1-M_2\right) \left(1-M_1\right) }{ \left(M_2-M_1^2\right)}$,
$I_x(a,b)$  is the regularized incomplete Beta function,
and $B(a,b)$ is the Beta function. The  beta distribution~\cite{Haenggi2016,Wang2017,Cui2017CoMP,Salehi2017} and the generalized beta distribution~\cite{Deng2017} have been shown to match  the exact meta distribution.

\item {\em Bounds on the meta distribution} are also presented in \cite[Corollary 4]{Haenggi2016}. For the Markov's bound, we can use any moment of $(1-P_{\rm s})$ and $P_{\rm s}$. For the Chebyshev's bound, we need the mean ($M_1$) and variance ($M_2-M_1^2$) of CSP. For the Paley-Zygmund (or Cauchy-Schwartz) bound, we simply need the first moment $M_1$.
\end{itemize}

For a given realization of transmitters $\Phi$, the transmission success events at a receiver {are obtained by averaging} over the fading channels and are thus i.i.d. over time. The {\em local delay} (defined as the number of transmission attempts until a packet is successfully received \cite{Baccelli2010,Haenggi2013}), is thus geometrically distributed with parameter $P_{\rm s}$. 
\begin{Definition}[Distribution of the Local Delay] {For a given realization}, {local delay}, $L$, follows a geometric  distribution with parameter $P_{\rm s}$ given in {\bf Definition~1}, {i.e.,}
\begin{IEEEeqnarray}{rCl}
	\mathbb{P}\left(L=k \mid \Phi \right) &=& \left(1-P_{\rm s}\right)^{k-1}P_{\rm s}, \qquad k\in\mathbb{N}.
\end{IEEEeqnarray}
{Therefore,} the mean local delay is given by $\mathbb{E}\left[ L \right] = \mathbb{E}_{\Phi}\left[ \mathbb{E}\left[ L \mid \Phi \right] \right]
	= \mathbb{E}_{\Phi}\left[ \frac{1}{P_{\rm s}} \right]
	= M_{-1}$ and the variance of the local delay is
$
\mathbb{E}\left[ L^2 \right]-\mathbb{E}\left[ L \right]^2 = \mathbb{E}_{\Phi}\left[ \mathbb{E}\left[ L^2 \mid \Phi \right] \right] - M_{-1}^2=  2M_{-2}-M_{-1}-M_{-1}^2,
$
\end{Definition}
For ergodic point processes, now we are able to answer the question ``What fraction of users  successfully receive their desired signals (i.e.,  SIR constraint satisfied) in at most $k$ time slots with probabilities larger than $x$?". We can answer this question by deriving the following:
\begin{IEEEeqnarray}{rCl}
	\mathbb{P}^{!0}\left( \mathbb{P}\left( L \le k \mid \Phi \right)>x \right) \stackrel{(a)}{=}  \mathbb{P}^{!0}\left( 1-\left(1-P_{\rm s}\right)^k > x \right) 
= \bar{F}_{P_{\rm s}}(1-\left(1-x\right)^{1/k}), 
	\label{distributionLD}
\end{IEEEeqnarray} 
where (a) is obtained by CDF of the geometric distribution, and $\bar{F}_{P_{\rm s}}(.)$ is the meta distribution defined in \eqref{MD}. Based on \eqref{distributionLD}, the meta distribution also reveals the distribution of the CSP for any number of retransmissions. 

\noindent
{\bf Example:} With $x=0.95$, $\bar{F}_{P_{\rm s}}(0.95)$ is the fraction of users that successfully receive their desired signals (or the SIR is higher than the target threshold) in the first transmission attempt (i.e., $k = 1$) with  a probability  higher than 0.95 (i.e., with reliability 0.95). $\bar{F}_{P_{\rm s}}(0.78)$ is the fraction of users that successfully receive their desired signals after the second transmission attempt (i.e., $k=2$) with reliability 0.95. $\bar{F}_{P_{\rm s}}(0.63)$ is the fraction of users that successfully receive their desired signals after the third transmission attempt (i.e., $k=3$) with reliability 0.95. 
$\bar{F}_{P_{\rm s}}(0.63)$ can also be interpreted as the fraction of users that successfully receive their desired signals in the first time slot with reliability 0.63, {or the fraction of users that successfully receive their desired signals after the second time slot with reliability 0.86.}

\section{System Model and Assumptions}
This section details the network model, channel model, and interference model along with assumptions for multi-user  uplink and downlink NOMA systems.
\subsection{Uplink NOMA}

\subsubsection{Network and Channel Model} We consider an uplink  NOMA system where BSs are distributed according to a homogeneous PPP\footnote{The motivation of modeling BS locations for real-world cellular networks with PPP was justified in \cite{Andrews2016}.} $\Phi_{\rm B}$ of intensity $\lambda_{\rm b}$. {Each user is connected to its nearest BS and there are at least $N$ users in each Voronoi cell\footnote{{This can be viewed as the general case of the {\em user point process of type I} introduced in \cite{Haenggi2017user}. In \cite{Haenggi2017user}, the user point process for $N=1$ is studied which is the case in OMA. In this paper, we consider $N\ge1$.}}. 
We consider random user selection, i.e.,  $N$ users are randomly selected for NOMA transmission from users  located in the Voronoi cell. The network is interference-limited.} The channel power between a user located at $x$ and the typical BS located at the origin is given by $h_x\ell (x)$ where $h_x$ represents the small-scale multi-path fading channel powers following i.i.d. exponential distribution with unit mean and $\ell(x)=\|x\|^{-\alpha}$ represents {the path-loss with exponent $\alpha$, where $\alpha>2$.} 

\subsubsection{SIC} We consider perfect SIC, i.e., the BS perfectly decodes and cancels the first $m-1$ strong interference signals before decoding the signal of the $m$-th rank user. 
The channel gains of different users are different\footnote{{The channel frequency/bandwidth is same for all users in NOMA; however, the channel gain experienced by the users on that specific frequency will be different due to their path-loss and fading.}} in the uplink; therefore, each message signal experiences distinct channel gain. The conventional uplink power control, which is typically intended to equalize the received signal powers of users, removes the channel distinctness and thus will not be feasible for uplink NOMA~\cite{Tabassum2017}. Therefore, in the uplink, we assume that all  users transmit with the same power~$P$.

\subsubsection{Interference and SIR Model} To model the intra-cell interference with SIC, first the  typical BS needs to  rank the received powers of various users. 
However, note that the impact of path-loss factor is more stable and  dominant compared to the instantaneous multi-path channel fading effects.  Therefore, the order statistics of the distance outweigh the fading effects, which vary on a much shorter time scale.
As such, the ranking of users in  terms of their distances from the serving BS is generally considered as a reasonable approximation of their respective ranked received signal powers~\cite{approx1,Tabassum2017}. This approximation provides tractability in the analysis. 
The intra-cell interference for the $m$-th rank user can therefore be modeled as:
\begin{IEEEeqnarray}{rCl}
	I_{(m)}^{\rm intra}=\sum_{i=m+1}^{N} P h_{x_{(i)}} \|{x_{(i)}}\|^{-\alpha}, \qquad m=1,2,...,N,
	\label{eq:I_intra_uplink}
\end{IEEEeqnarray}
where $h_{x_{(i)}}$ is the fading from the $i$-th rank user located at $x_{(i)}$ (in the Voronoi cell of the typical BS) to the typical BS. The inter-cell interference {is} given as follows:
\begin{IEEEeqnarray}{rCl}
	I^{\rm inter}=\sum_{x \in \Phi_{\rm I}} P h_x \|x\|^{-\alpha},
	\label{eq:I_inter_uplink}
\end{IEEEeqnarray}
where $\Phi_{\rm I}$ is the  point process describing the locations of the inter-cell interferers, which is unknown. Using \eqref{eq:I_intra_uplink} and \eqref{eq:I_inter_uplink}, for the user at rank $m$, the SIR is given as follows:
\begin{IEEEeqnarray}{rCl}
	\SIR_{(m)}=\frac{ P h_{x_{(m)}} \|{x_{(m)}}\|^{-\alpha} }{I_{(m)}^{\rm intra} + I^{\rm inter}}.
	\label{eq:SIR_uplink}
\end{IEEEeqnarray}

\subsection{Downlink NOMA}

\subsubsection{Network and Channel Model} Similar to uplink, we consider downlink  NOMA system with $N$ users in {each} Voronoi cell. BSs are distributed according to the homogeneous PPP $\Phi_{\rm B}$ of intensity $\lambda_{\rm b}$ and each BS can transmit with maximum power $P$.  The effect of thermal noise is neglected.  The channel power gain between the BS located at $x$ and the typical user located at the origin is given by $h_x\ell (x)$, and $h_x$ for different BSs are modeled by i.i.d. exponential random variables with unit mean. $\ell(x)=\|x\|^{-\alpha}$  represents the power-law path-loss, in which $\alpha>2$ is the path-loss exponent.
The power allocated to the $i$-th rank user is $P_i=\beta_i P$, $\forall i=1,2,\cdots,N$. Also, we have $\beta_i\le \beta_j$, $\forall i\le j$ such that $\sum_{i=1}^{N} \beta_i=1$ (or equivalently, $\sum_{i=1}^{N} P_i=P$).

\subsubsection{SIC} We consider perfect SIC, i.e., user at rank $m$ successfully removes the intra-cell interference of all users who are at higher ranks in terms of their distances. 
\subsubsection{Interference and SIR Model} The intra-cell interference at $m$-th rank user can be given as: 
\begin{IEEEeqnarray}{rCl}
	I_{(m)}^{\rm intra}=\sum_{i=1}^{m-1} \beta_i P h_0 \|x_0\|^{-\alpha}, \qquad m=1,2,\cdots,N,
	\label{eq:I_intra_downlink}
\end{IEEEeqnarray}
where $h_0$ is the fading from the serving BS located at $x_0$ to the $m$-th rank user located at the origin. 
The inter-cell interference can be modeled as:
\begin{IEEEeqnarray}{rCl}
	I_{(m)}^{\rm inter}=\sum_{x \in \Phi_{\rm B} \setminus \{x_0\}} P h_x \|x\|^{-\alpha}, \qquad m=1,2,\cdots,N.
	\label{eq:I_inter_downlink}
\end{IEEEeqnarray}
Hence, for the user at rank $m$, the SIR can be given as:
\begin{IEEEeqnarray}{rCl}
	\SIR_{(m)}=\frac{ \beta_m P h_0 \|x_0\|^{-\alpha} }{I_{(m)}^{\rm intra} + I_{(m)}^{\rm inter}}.
	\label{eq:SIR_downlink}
\end{IEEEeqnarray}
By Slivnyak's theorem, the point process for the inter-cell interferers is a PPP  with intensity $\lambda_{\rm b}$ in $\mathbb{R}^2\setminus b(o,\|x_0\|)$ where distribution of $\|x_0\|$ depends on the rank of the user. This is different from uplink where the inter-cell interference is received at a typical BS and is therefore same for all NOMA users in the typical Voronoi cell.
\section{Uplink NOMA: Moments and Meta Distribution of the CSP}
In this section, we derive the  CSP, and the moments and the meta distribution of the CSP for an uplink NOMA network. For this,
\begin{itemize}
\item we first derive the distance distribution of the intra-cell interferers,
\item then we derive approximate point process of the inter-cell interferers,
\item and  then we derive the moments of CSP. The exact and approximate meta distributions can then be obtained using \eqref{exactMD} and \eqref{approxMD}, respectively, as described in Section~II.
\end{itemize}
For performance analysis of uplink NOMA, modeling the actual point process $\Phi_{\rm I}$ for the inter-cell interferers is critical.
In this section,  we propose two point processes to approximate $\Phi_{\rm I}$. We will demonstrate the accuracy of the proposed point processes by comparing the second moment measure of each point process with the second moment measure of the actual (or original) point process $\Phi_{\rm I}$ obtained by simulations. Since the typical BS is located at the $o$ (origin) and we model the interferers' point process from the perspective of the typical BS, we are interested in the first and second moment measures for $b(o,r)$, where $b(o,r)$ denotes the ball of radius $r$ centred at $o$. 

\subsection{Distance Distributions of the Intra-cell Interferers} \label{subsec:distance_distribution_uplink}
For any uplink user in the typical Voronoi cell, the probability density function (PDF) and the cumulative density function (CDF) of the desired link distance are given as follows \cite{Wang2017}:
\begin{IEEEeqnarray}{rClr}
f_R (r)= (5/2) \lambda_{\rm b} \pi r e^{-(5/4)\lambda_{\rm b} \pi r^2}, \qquad
F_R (r)= 1 - e^{-(5/4)\lambda_{\rm b} \pi r^2}.            \label{eq:CDFdistance_uplink}
\end{IEEEeqnarray}
Note that the distance of the user to the typical BS {is} not Rayleigh distributed {with mean $1/( 2\sqrt{\lambda_{\rm b}} )$}~\cite{Haenggi2017user}. 
Using the above distributions and order statistics, the distribution of the distance of the user at rank $m$ from its serving BS can be derived as follows \cite{david2003order,yang2011order}:
\begin{IEEEeqnarray}{rCl}
	f_{R_m} (r)&=& \frac{5 \lambda_{\rm b} \pi r \left( 1 - e^{-(5/4) \lambda_{\rm b} \pi r^2} \right)^{m-1} \left( e^{-(5/4) \lambda_{\rm b} \pi r^2} \right)^{N-m+1}}{2 B(N-m+1,m)}, \qquad r \ge 0,
	\label{eq:PDFdistance_m_uplink}
\end{IEEEeqnarray}
where $B(\cdot,\cdot)$ is the Beta function.
Conditioned on the distance of the user at rank $m$ ($R_m=r_m$), it was shown in \cite{Afshang2016,Tabassum2017} that the distances of users at lower or higher ranks than the $m$-th rank user to the typical BS are i.i.d. and their PDFs can be characterized, respectively, as follows: 
\begin{align}
	f_{R_{\rm in}}\left(r \mid R_m=r_m \right)= \frac{f_R (r)}{F_R (r_m)}, \qquad r\le r_m, \quad i=1,\cdots,m-1,
	\\
	f_{R_{\rm out}}\left(r \mid R_m=r_m \right)= \frac{f_R (r)}{1-F_R (r_m)}, \qquad r\ge r_m, \quad i=m+1,\cdots,N.
	\label{eq:PDFdistance_out_uplink}
\end{align}

\subsection{BS/user Pair Correlation Function}

{In this subsection, we  obtain the pair correlation function of $\Phi_{\rm I}$ with respect to the origin (the location of the typical BS) through simulations, and then, to model the locations of the interferers, propose two point processes with the same BS/users pair correlation function.}

\begin{Definition}[BS/user Pair Correlation Function~\cite{Haenggi2017user}]
For the BSs point process $\Phi_{\rm B}$ of intensity $\lambda_{\rm b}$, the BS/user pair correlation function $g_{\lambda_{\rm b}}(r)$ can be defined as follows:
\begin{IEEEeqnarray}{rCl}
	g_{\lambda_{\rm b}}(r) \triangleq 
	\frac{1}{2 \pi r} \frac{{\rm d}}{{\rm d}r} K(r) =
	\frac{1}{2 \pi r} \frac{{\rm d}}{{\rm d}r} \left(\frac{1}{N\lambda_{\rm b}} \mathbb{E}^0 [\Phi_{\rm I}(b(o,r))] \right),
	\label{eq:pcf}
\end{IEEEeqnarray}
where $\mathbb{E}^0$ is the Palm expectation (given that the typical BS is at the origin). {When $\Phi_{\rm I}$ is scale-invariant, $g_{\lambda_{\rm b}}(r)=g_1(\sqrt{\lambda_{\rm b}}r)$.}
\end{Definition}
Note that the BS/user pair correlation function $g_{\lambda_{\rm b}}(r)$ is useful in  approximating the interfering users' point process by a PPP of intensity function $\lambda_{\rm b}g_{\lambda_{\rm b}}(r)$ \cite{Haenggi2017user}. Specifically, \cite{Haenggi2017user}  studied the point process of uplink interferers for $N=1$ (i.e., for orthogonal multiple access [OMA]), and through numerical fitting, the best exponential fit for $N=1$ was obtained as follows:
\begin{IEEEeqnarray}{rCl}
	g_1(r)=1-e^{-(12/5) \pi r^2}.
	\label{eq:best_exponential_fit}
\end{IEEEeqnarray}
Note that any other point process with the same intensity function ($\lambda_{\rm b}g_{\lambda_{\rm b}}(r)$) can also be used to approximate the interfering users' point process. 

\begin{figure}[t]
	\centering
	\includegraphics[width=\figwidth]{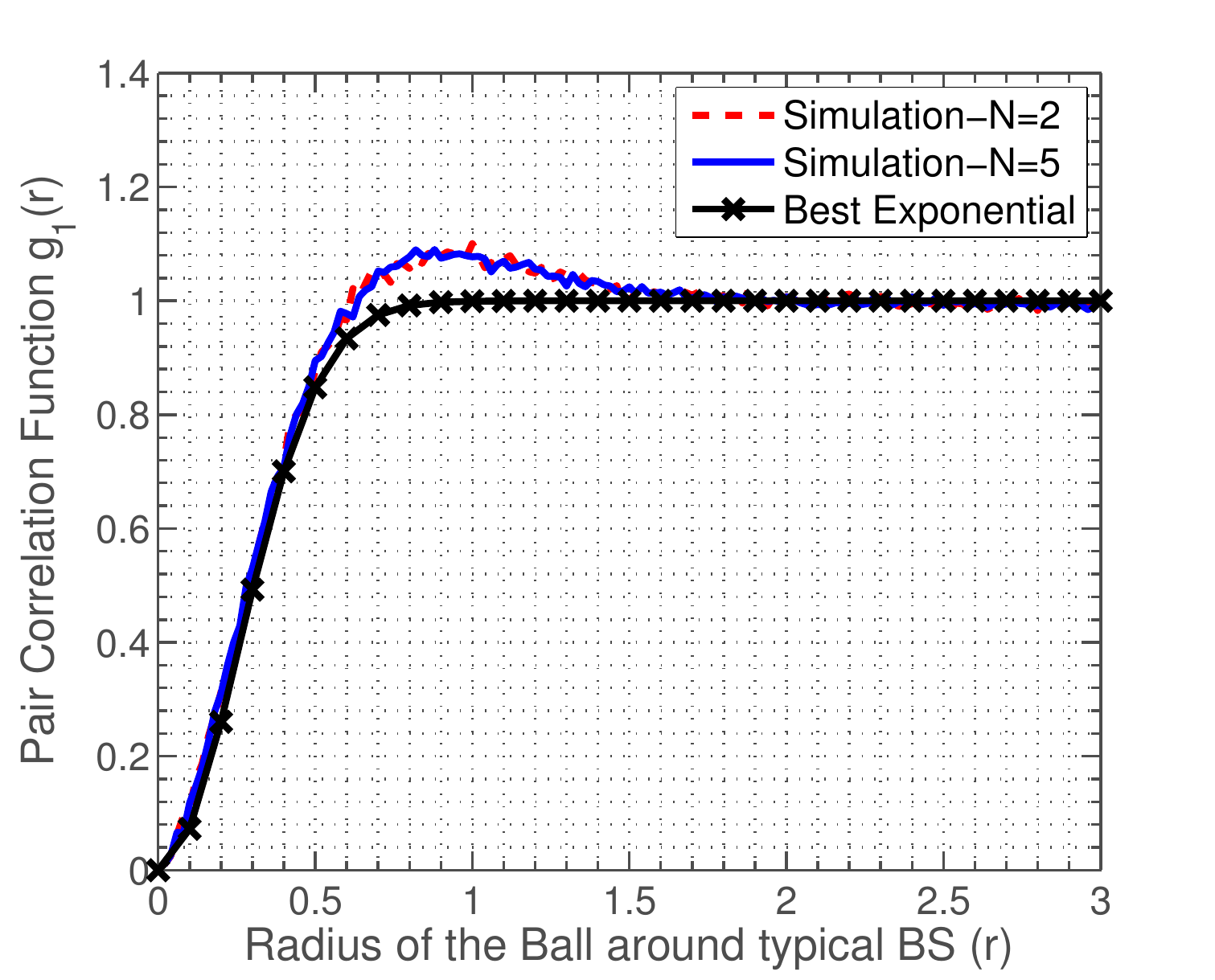}
	\caption{BS/user pair correlation function $g_1(r)$ for different $N$ and validation with  the best exponential fit in \eqref{eq:best_exponential_fit} for $N$=1.}
	\label{fig:pcf}	
\end{figure}

Along the same lines, we also obtain $g_{1}(r)$ through simulations.  In \figref{fig:pcf}, $g_1(r)$ is illustrated for $N=2$ and $N=5$,  and we observe that it does not vary for different values of $N$. {The reason is that  the average number of inter-cell interferers within the distance $r$ from the typical BS, $\mathbb{E}^0 [\Phi_{\rm I}(b(o,r))]$, for clusters of $N$ users in  NOMA is $N$ times higher than that in OMA. Therefore, \eqref{eq:pcf} does not change with respect to $N$}. We also compare the simulation results with the best exponential fit for $N=1$. 
Using the invariance property of $g_1(r)$ with respect to $N$ along with the {the scale-invariance property of the model} and the results in \cite{Haenggi2017user}, we  approximate the inter-cell interferers' point process  $\Phi_{\rm I}$ by a PPP $\dbar{\Phi}_{\rm I}$ with intensity $N\lambda_{\rm b}g_{\lambda_{\rm b}}(r)$. {In each NOMA cluster, users are located close to each other in the same Voronoi cell; however,} 
the points of the PPP are independent from each other~\cite{haenggi2012stochastic}. Therefore, approximating the inter-cell interferers with $\dbar{\Phi}_{\rm I}$ may not capture the dependence of the inter-cell interferers' locations in each NOMA cluster precisely. To address this issue, we also propose a cluster process $\bar{\Phi}_{\rm I}$ to approximate  $\Phi_{\rm I}$. In the following, first we define the intensity measure and then we describe the two proposed models along with their validation and comparative analysis.

{
	\begin{Definition} [Intensity Measure~\cite{haenggi2012stochastic}] 
	For any point process $\Phi$, the intensity measure (first moment measure) $\Lambda(B)$ is the mean number of points in $B$, i.e.,
$
		\Lambda(B)=\mathbb{E} \Phi(B), \quad \forall B\subset\mathbb{R}^2.
$
If $\Phi$ has an intensity function $\lambda(x)$, then $\Lambda(B)=\int\limits_B \lambda(x) {\rm d}x$.
\end{Definition}

\subsection{Interferers' Point Process Models}
\subsubsection{Model 1} To model the interferers, we consider a PCP ${\bar{\Phi}}_{\rm I}$\footnote{{If the parents of a cluster process are the points of a Poisson process, the resulting process is a Poisson cluster process (PCP) \cite{haenggi2012stochastic}.}}, where the parents form an inhomogeneous PPP $\bar{\Phi}_{\rm P}$ with intensity function $\bar{\lambda}_{\rm p}(x)=\lambda_{\rm b} \left( 1-e^{-(12/5) \lambda_{\rm b} \pi \|x\|^2} \right)$. In each cluster, $N$ offspring points are located in the same location as the parent, i.e., {for a parent at $x$}, $N$ offsprings are i.i.d. with PDF $f(y)=\delta(y-x)$, where $x,y\in\mathbb{R}^2$. {This model can also be viewed as a non-simple PPP~\cite{haenggi2012stochastic}. As mentioned earlier, other cluster processes that have the same BS/user pair correlation function can also be used to model inter-cell interferers, but $\bar{\Phi}_{\rm I}$ is more tractable.} 

Using \textbf{Model 1}, the mean number of inter-cell interferers within the distance $r$ from the typical BS (first moment of ${\bar{\Phi}}_{\rm I}(b(o,r))$) can be derived as follows: 
\begin{IEEEeqnarray}{rCl}
	\bar{\Lambda}(b(o,r)) &=& \mathbb{E} \left[ {\bar{\Phi}}_{\rm I}(b(o,r)) \right]   \overset{\text{(a)}}{=}  N\mathbb{E} \left[ {\bar{\Phi}}_{\rm P}(b(o,r)) \right] = N \bar{\Lambda}_{\rm p}(b(o,r)) \nonumber
\\
	&=& N \int_{b(o,r)} \bar{\lambda}_{\rm p}(x) {\rm d}x=
  N \lambda_{\rm b} \left[ \pi r^2 - \frac{5}{12 \lambda_{\rm b}} \left( 1-e^{-(12/5) \lambda_{\rm b} \pi r^2} \right) \right],
	\label{eq:first_moment_for_type_I}
\end{IEEEeqnarray}
where $\bar{\Lambda}$ and $\bar{\Lambda}_{\rm p}$ are the intensity measures of ${\bar{\Phi}}_{\rm I}$ and ${\bar{\Phi}}_{\rm P}$, respectively, and step (a) follows from ${\bar{\Phi}}_{\rm I}(b(o,r))=N{\bar{\Phi}}_{\rm P}(b(o,r))$. The second moment measure of ${\bar{\Phi}}_{\rm I}(b(o,r))$ is derived as follows:
\begin{IEEEeqnarray}{rCl}
	\mathbb{E} \left[ {\bar{\Phi}}_{\rm I}^2(b(o,r)) \right]  &=& N^2 \mathbb{E} \left[ {\bar{\Phi}}_{\rm P}^2(b(o,r)) \right] \overset{\text{(a)}}{=} N^2 \sum_{k=0}^{\infty} k^2 \frac{ \bar{\Lambda}_{\rm p}(b(o,r))^k }{ k! } e^{-\bar{\Lambda}_{\rm p}(b(o,r))}
\nonumber\\
	 &\overset{\text{(b)}}{=}&  N^2 \left[ \bar{\Lambda}_{\rm p}(b(o,r)) + \bar{\Lambda}_{\rm p}(b(o,r))^2 \right]
	 \overset{\text{(c)}}{=}  \bar{\Lambda}(b(o,r)) \left[ N +  \bar{\Lambda}(b(o,r)) \right], \quad
	\label{eq:second_moment_for_type_I}
\end{IEEEeqnarray}
where (a) follows since ${\bar{\Phi}}_{\rm P}(b(o,r))$ is a Poisson random variable with mean $\bar{\Lambda}_{\rm p}(b(o,r))$, (b) is obtained from mean and variance of the Poisson distribution, and (c) follows by 
$\bar{\Lambda}(b(o,r))=N \bar{\Lambda}_{\rm p}(b(o,r))$, {where $\bar{\Lambda}(b(o,r))$} is given in \eqref{eq:first_moment_for_type_I}. \eqref{eq:second_moment_for_type_I} can also be derived using the second factorial moment measure of PPPs.

\subsubsection{Model 2} In this model, we approximate $\Phi_{\rm I}$ with an inhomogeneous PPP ${\dbar{\Phi}}_{\rm I}$ with intensity function $\dbar{\lambda}(x)=N \lambda_{\rm b} \left( 1-e^{-(12/5) \lambda_{\rm b} \pi \|x\|^2} \right)$. The mean number of inter-cell interferers within the distance $r$ from the typical BS (first moment of ${\dbar{\Phi}}_{\rm I}(b(o,r))$) {is} as follows: 
\begin{IEEEeqnarray}{rCl}
	\dbar{\Lambda}(b(o,r)) 
	= \mathbb{E} \left[ {\dbar{\Phi}}_{\rm I}(b(o,r)) \right]
	= \int_{b(o,r)} \dbar{\lambda}(x) {\rm d}(x)  
	= \bar{\Lambda}(b(o,r)), \nonumber
	\label{eq:first_moment_for_type_II}
\end{IEEEeqnarray}
where $\dbar{\Lambda}$ denotes the intensity measure of ${\dbar{\Phi}}_{\rm I}$. The second moment of ${\dbar{\Phi}}_{\rm I}(b(o,r))$ is  given by
\begin{IEEEeqnarray}{rCl}
    \mathbb{E} \left[ {\dbar{\Phi}}_{\rm I}^2(b(o,r)) \right]  
    &\overset{\text{(a)}}{=}& \dbar{\Lambda}(b(o,r)) \left[\dbar{\Lambda}(b(o,r))+1\right], 
	\label{eq:second_moment_for_type_II}
\end{IEEEeqnarray}
where (a) follows since ${\dbar{\Phi}}_{\rm I}(b(o,r)$ is  Poisson variable with mean $\dbar{\Lambda}(b(o,r))$.}
Note that {the point process introduced in \cite{Haenggi2017user} is a special case of the {proposed models 1 and 2} when $N=1$.}

\subsection{Model Validation}
\begin{figure}
	\centering
	\includegraphics[width=\figwidth]{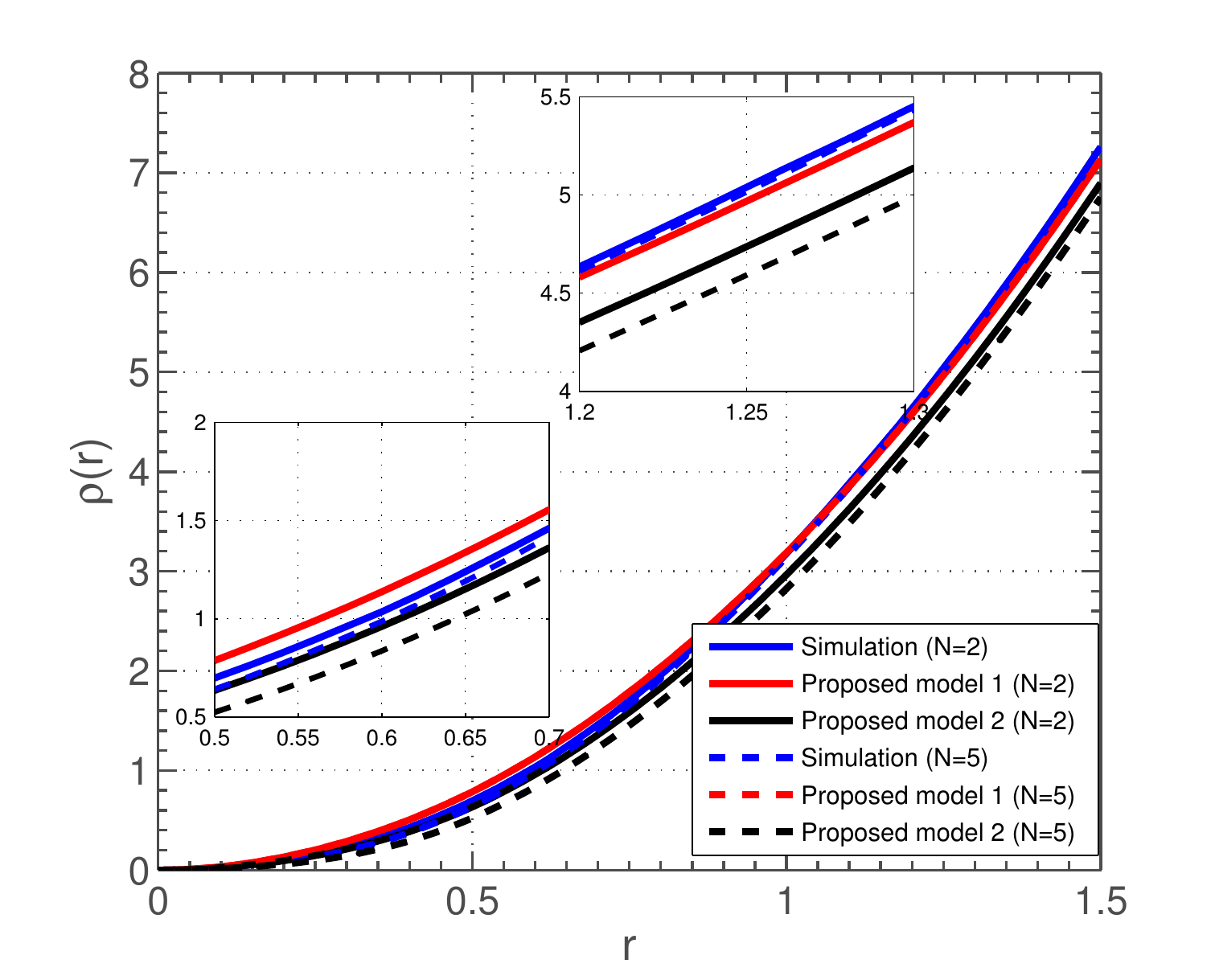}
	\caption{Comparison of the second moment measure of $\Phi_{\rm I}$ with those of the proposed models (derived in \eqref{eq:second_moment_for_type_I} and \eqref{eq:second_moment_for_type_II}) for $\lambda_{\rm b}=1$.}
	\label{fig:second_moment}	
\end{figure}

To compare the second moment of $\Phi_{\rm I}(b(o,r))$ with the proposed models, we define
$
\rho(r)\triangleq  \frac{1}{ N \lambda_{\rm b} } \sqrt{ \mathbb{E}\left[ {\Phi}_{\rm I}^2(b(o,r)) \right] } 
$. We consider the square root of the normalized second moment since it illustrates the difference between the models better. In \figref{fig:second_moment}, $\rho(r)$ for the original interferers point process $\Phi_{\rm I}$ is obtained via simulations and a comparison is performed with the proposed models. We observe that, $\rho(r)$ for the proposed models {are} close to the $\rho(r)$ of $\Phi_{\rm I}$. Moreover, based on \figref{fig:second_moment}, \textbf{Model 1} provides a better approximation for larger values of $r$.

\subsection{Moments and Meta Distribution of the CSP $(P_s(\theta))$}
The moments of the CSP for uplink NOMA users can be derived as follows.
\begin{theorem} [Moments of the CSP for Uplink NOMA] In  uplink NOMA, $b$-th moment of the CSP, $b\in\mathbb{C}$,  for the $m$-th rank user can be derived as follows:
\begin{IEEEeqnarray}{rCl}
		M_{b,(m)}=
	    \int_{0}^{\infty} \left[ (5/2) \lambda_{\rm b} \pi r^2 e^{(5/4) \lambda_{\rm b} \pi r^2} \mu_{b} \left( (5/4) \lambda_{\rm b} \pi r^2, \theta \right) \right]^{N-m}
		\mathbb{E} 
		\left[ \prod_{x\in\Phi_{\rm I}} \left( \frac{1}{ 1+\theta r^{\alpha}\|x\|^{-\alpha} } \right)^b  \right]  
		f_{R_m}(r) {\rm d}r, \nonumber
		\\
		\label{eq:theorem_moment_CSP_uplink}
    \end{IEEEeqnarray}
    where $f_{R_m}(r)$ is given in \eqref{eq:PDFdistance_m_uplink} and
    $\mu_b(x,z) = \int_0^1 \frac{ t^{-3}e^{-x t^{-2}} }{\left( 1+z t^{\alpha} \right)^b} {\rm d}t.$ The expectation in \eqref{eq:theorem_moment_CSP_uplink}, which is conditioned on the serving distance $r$ can be approximated, using the proposed \textbf{ Model 1} and \textbf{Model 2} for inter-cell interferers' point process, respectively, as follows:
    \begin{IEEEeqnarray}{rCl}
        \Scale[1.04]{\mathbb{E} \left[ \prod\limits_{x\in\Phi_{\rm I}} \left( \frac{1}{ 1+\theta r^{\alpha}\|x\|^{-\alpha} } \right)^b \right] \approx
         \exp \left\{-2 \pi \lambda_{\rm b} \int\limits_0^{\infty} 
        \left[ 1-\left( \frac{1}{ 1+\theta r^{\alpha}x^{-\alpha} } \right)^{Nb} \right] 
        \left( 1- e^{-(12/5)\lambda_{\rm b}\pi x^2}\right) x {\rm d}x \right\}}, \IEEEeqnarraynumspace
\label{eq:inter_cell_interference_uplink_type_I}
    \\    
     \Scale[1.04]{\mathbb{E} \left[ \prod\limits_{x\in\Phi_{\rm I}} \left( \frac{1}{ 1+\theta r^{\alpha}\|x\|^{-\alpha} } \right)^b \right]\approx \exp \left\{-2 \pi N \lambda_{\rm b} \int\limits_0^{\infty} 
    	\left[ 1-\left( \frac{1}{ 1+\theta r^{\alpha}x^{-\alpha} } \right)^{b} \right] \left( 1- e^{-(12/5)\lambda_{\rm b}\pi x^2}\right) x {\rm d}x \right\}}. \IEEEeqnarraynumspace
    	\label{eq:inter_cell_interference_uplink_type_II}
    \end{IEEEeqnarray}
\label{Thm1}
\end{theorem}
\begin{IEEEproof} 
See  {\bf Appendix A}.
\end{IEEEproof}
The proposed  point processes  provide better approximations for standard transmission success probability {($b=1$)} when the SIR threshold is low as shown below.
{
\begin{Corollary}
{For $b=1$ (standard success probability), the proposed point process models provide better approximations when $\theta\to 0$.}
\end{Corollary}
\begin{IEEEproof} For $b=1$, we have 
$
	\mathbb{E} \left[ \prod_{x\in\Phi_{\rm I}}  \frac{1}{ 1+\theta r^{\alpha}\|x\|^{-\alpha} }   \right] 
	= L_{ I^{\rm{inter}} }(\theta r^{\alpha}/P),
$
    where $I^{\rm inter}$ is given in \eqref{eq:I_inter_uplink} and $L_{ I^{\rm{inter}} }(s)=\mathbb{E}\left[ e^{-sI^{\rm{inter}}} \right]$ is the Laplace transform of the inter-cell interference. When $s\to0$, we have
    \begin{IEEEeqnarray}{rCl}
    	L_{ I^{\rm{inter}} }(s)=\mathbb{E}\left[ e^{-sI^{\rm{inter}}} \right]
        \sim 1-\mathbb{E} \left[ sI^{\rm{inter}} \right] 
    	\overset{\text{(a)}}{=} 1-\mathbb{E} \left[ s\sum_{x\in\Phi_{\rm I}} Ph_x\|x\|^{-\alpha} \right]  
    	\overset{\text{(b)}}{=} 1- sP \mathbb{E} \left[ \sum_{x\in\Phi_{\rm I}} \|x\|^{-\alpha} \right], \nonumber
    \end{IEEEeqnarray}
    where (a) follows from \eqref{eq:I_inter_uplink}, and (b) follows since fading coefficients $h_x$ are i.i.d. with unit mean.
    
According to the Campbell's theorem, approximating $\Phi_{\rm I}$ with point processes that have the same BS/user pair correlation function (which can also be interpreted as the same intensity measure with respect to the origin) for any $f:\mathbb{R}^2\mapsto\mathbb{R}^+$ yields
    \begin{IEEEeqnarray}{rCl}
    	\mathbb{E} \left[ \sum_{x\in\Phi_{\rm I}} f(x) \right] \equiv 
    	\mathbb{E} \left[ \sum_{ x\in{\bar{\Phi}}_{\rm I} } f(x) \right] \equiv
    	\mathbb{E} \left[ \sum_{ x\in{\dbar{\Phi}}_{\rm I} } f(x) \right]. \nonumber
    \end{IEEEeqnarray} 
    Therefore, the proposed models provide better approximations for the first moment $M_1$ when $\theta\to 0$.
\end{IEEEproof}
Similarly, we can prove that for $m=1$ or larger values of $\lambda_{\rm b}$, the approximations are better, because the probabilities of small values of $r$ are higher for $m=1$ or larger values of $\lambda_{\rm b}$.

\begin{Corollary} \label{cor2}
	For $b\in\mathbb{R}$, $M_{b,(m)}$ of \textbf{Model 2} is a lower bound for $M_{b,(m)}$ of \textbf{Model 1}.
\end{Corollary}
\begin{IEEEproof} Using the identity 
	$1-y^N\equiv \left( 1-y \right) \left( 1+y+y^2+...+y^{N-1} \right)$, for $0 \le y$, we have, $1-y^N \le N(1-y)$.
	Then \textbf{Corollary \ref{cor2}} is obtained by setting $y=\left( \frac{1}{ 1+\theta r^{\alpha}x^{-\alpha} } \right)^{b}$ in \eqref{eq:inter_cell_interference_uplink_type_I} and \eqref{eq:inter_cell_interference_uplink_type_II}.
\end{IEEEproof}
}
The exact and approximate meta distributions of CSP can be obtained by using \eqref{exactMD} and \eqref{approxMD}, respectively, as described in Section~II.

\section{Downlink NOMA: Moments and Meta Distribution of the CSP}
In this section, we derive the  CSP, and the moments and meta distribution of the CSP in a downlink NOMA network. For this,  we first derive the distance distribution of the {desired link} and then derive the moments of CSP as well as the meta distribution. 

\subsection{{Distribution of the Desired Link Distance}}
Since each user connects to its nearest BS, the serving link distance distribution can be given by the Rayleigh distribution as follows \cite{Haenggi05tit}:
\begin{IEEEeqnarray}{rClr}
	f_R (r)= 2 \lambda_{\rm b} \pi r e^{-\lambda_{\rm b} \pi r^2},	\quad
	F_R (r)= 1 - e^{-\lambda_{\rm b} \pi r^2}.             \label{eq:CDFdistance_downlink}
\end{IEEEeqnarray}
Using the above equations and order statistics, the distribution of the distance of a user at rank $m$ from its serving BS can be given as follows \cite{david2003order,yang2011order}:
\begin{IEEEeqnarray}{rCl}
	f_{R_m} (r)&=& \frac{ 2 \lambda_{\rm b} \pi r  \left( 1 - e^{-\lambda_{\rm b} \pi r^2} \right)^{m-1} \left( e^{-\lambda_{\rm b} \pi r^2} \right)^{N-m+1}}{B(N-m+1,m)}, \qquad r \ge 0.
	\label{eq:PDFdistance_m_downlink}
\end{IEEEeqnarray}

\subsection{Moments and Meta Distribution of the CSP ($P_s(\theta)$)}
The $b$-th moment of the conditional success probability $M_{b,(m)}$, $b\in\mathbb{C}$, for an $m$-th rank downlink NOMA user is derived in the following. Based on these moments, we can derive the mean success probability, the meta distribution, and the mean local delay.

\begin{theorem}[Moments of the CSP for Downlink NOMA]  For a user at rank $m$, the $b$-th moment of the conditional success probability $M_{b,(m)}$ is  
{\begin{IEEEeqnarray}{rCl}
			M_{b,(m)}=
\begin{cases}
\frac{B(A_{b,m}+N-m+1,m)}{B(N-m+1,m)}, &  \theta<{\beta_m}/{\sum\limits_{i-1}^{m-1}\beta_i}\\
0,     & \theta\ge{\beta_m}/{\sum\limits_{i-1}^{m-1}\beta_i} \quad \& \quad \Re(b)>0\\
\infty,& \theta\ge{\beta_m}/{\sum\limits_{i-1}^{m-1}\beta_i} \quad \& \quad \Re(b)<0
\end{cases}
\label{eq:moment_CSP_downlink}
\end{IEEEeqnarray}}
where $A_{b,m}=\sum_{k=1}^{\infty} \binom{b}{k} (-1)^{k+1} c_m^k \frac{\delta}{k-\delta} \,_2F_1(k,k-\delta;k-\delta+1;-c_m)$, $c_m=( \frac{\beta_m}{\theta}-\sum\limits_{i-1}^{m-1}\beta_i )^{-1}$, $\delta=2/\alpha$, $_2F_1$ is the Gauss Hypergeometric function, {and $\Re(b)$ gives the real part of $b$}. 
		\label{Thm2}
\end{theorem}
\begin{IEEEproof} See {\bf Appendix B}.
\end{IEEEproof}
Note that the condition $\theta<{\beta_m}/\sum\limits_{i-1}^{m-1}\beta_i$ (or equivalently, $ 0< c_m < \infty $) implies that  the received SIR at the $m$-th rank user is greater than the required SIR $\theta$ in the absence of inter-cell interference. Moreover, when $N=1$, which is the case in orthogonal multiple access, {\bf Theorem \ref{Thm2}} reverts back to the known results for downlink Poisson cellular networks \cite{Haenggi2016}. 

In the following, a simplified closed-form expression for negative moments $M_{-w,(m)}$, $w\in\mathbb{R}^+$, is provided. The expression is useful in evaluating the mean local delay of an $m$-th rank user in closed-form by setting $w=1$. 
\begin{Corollary} When $b=-w$, $w\in\mathbb{R}^+$, {and $c_m>0$}
	{\begin{IEEEeqnarray}{rCl}
		M_{-w,(m)} =
		\begin{cases}
			\frac{B(N-m-D_{w,m}+1,m)}{B(N-m+1,m)}, & D_{w,m} < N-m+1 \\
			\infty, & \rm{otherwise}
		\end{cases}
	\end{IEEEeqnarray}}
	where $D_{w,m}=\sum\limits_{k=1}^{\infty} \binom wk c_m^k \frac{\delta}{k-\delta}$. {When $c_m<0$, from {\bf Theorem \ref{Thm2}}, we have $M_{-w,(m)}=\infty$.}  
	\label{corollary1}
\end{Corollary}
\begin{IEEEproof}
		From the proof of the {\bf Theorem \ref{Thm2}}, we have
		\begin{IEEEeqnarray}{rCl}
			M_{-w,(m)}  &=& \mathbb{E}_{R_m} \left[ \exp\left\{ -2\pi\lambda_{\rm b} \int\limits_{R_m}^{\infty} 
			\left[ 1-\left( \frac{1}{1+c_mR_m^{\alpha}r^{-\alpha}} \right)^{-w} \right] r {\rm d}r \right\} \right] \nonumber
			\\
			&\stackrel{(a)}{=}& \int\limits_{0}^{\infty} \exp \left\{ \pi \lambda_{\rm b} r^2 \sum_{k=1}^{\infty} \binom wk c_m^k \frac{\delta}{k-\delta}  \right\} f_{R_m}(r) {\rm d}r.
			\label{eq:cor. proof}
		\end{IEEEeqnarray}
		Finally, \textbf{Corollary \ref{corollary1}} is obtained by substituting \eqref{eq:PDFdistance_m_downlink} in \eqref{eq:cor. proof} and setting $D_{w,m}=\sum_{k=1}^{\infty} \binom wk c_m^k \frac{\delta}{k-\delta}$.
\end{IEEEproof}
Note that setting $b=jt$, $j=\sqrt{-1}$ and $t\in\mathbb{R}^+$, the exact meta distribution of the $m$-th rank user is derived by substituting $M_{jt,(m)}$ from {\bf Theorem \ref{Thm2}} in \eqref{exactMD}. 

Since the exact meta distribution is complicated and does not provide any direct insights, the corresponding beta approximation  is defined. To derive the beta approximation, we need the first and second moments of $P_{{\rm s},(m)}$. The standard (mean) success probability, which is the first moment of $P_{{\rm s},(m)}$, can be easily obtained by setting $b=1$ in {\bf Theorem \ref{Thm2}}, i.e.,  $M_{1,(m)}={B(A_{1,m}+N-m+1,m)}/{B(N-m+1,m)}$, where 
$A_{1,m}=c_m \frac{\delta}{1-\delta} \, _2F_1(1,1-\delta;2-\delta;-c_m)$. Similarly, we can derive the second moment $M_{2,(m)}$ by setting $b=2$. The beta approximation is obtained by substituting $M_{1,(m)}$ and $M_{2,(m)}$ in \eqref{approxMD} as described in Section~II.

\begin{Corollary}[Local Delay of User at Rank $m$]
For the $m$-th rank user, when {$c_m>0$ and} $D_{1,m}=c_m \frac{\delta}{1-\delta} < N-m+1$, the mean local delay is finite and is given by
\begin{equation}
M_{-1,(m)}=\frac{B(N-m-D_{1,m}+1,m)}{B(N-m+1,m)}.
\end{equation} 
When {$c_m<0$ or} $D_{1,m}=c_m \frac{\delta}{1-\delta}\ge N-m+1$ , the mean local delay  is infinite.
\end{Corollary}

\section{Application of the Analytical Framework}
In this section, we demonstrate one application of the developed analytical framework for  optimal transmit power allocation in a large-scale downlink NOMA network with an objective to maximizing the standard success probability of  a given user. We first consider a two-user NOMA system for which closed-form solutions are obtained and then we consider an $N$-user NOMA system for which the solutions can be obtained numerically.

\subsection{Transmit Power Optimization in Two-User Downlink NOMA}
 For a two-user NOMA system,   we maximize the success probability of user at 2-nd rank  $M_{1,(2)}$ with constraints on the minimum success probability achieved by the 1-st rank user  $M_{1,(1)}$ in order to optimize the power allocation coefficients  of users $\beta_1, \beta_2$. This optimization problem is referred to as {\bf P1}. We further extend the optimization problem {\bf P1} by incorporating the constraints on mean local delays for each user and refer to the extended optimization problem as {\bf P2}. For the two-user NOMA system, we use the closed-form solutions obtained in  the previous section.

Given $N=2$, $c_1=\left( \beta_1/\theta \right)^{-1}>0$, and $c_2=\left( \beta_2/\theta - \beta_1 \right)^{-1}>0$, we obtain the average CSP of the 1-st and 2-nd rank users, respectively, as follows:
\begin{align*}
M_{1,(1)}=\frac{2}{2+A_{1,1}},\quad
M_{1,(2)}=\frac{2}{  2+3 A_{1,2}+A^2_{1,2} },
\end{align*}
where $A_{1,1}=c_1 \frac{\delta}{1-\delta}\,_2F_1(1,1-\delta,2-\delta,-c_1)$ and $A_{1,2}=c_2 \frac{\delta}{1-\delta}\,_2F_1(1,1-\delta,2-\delta,-c_2)$. As mentioned in {\bf Theorem \ref{Thm2}}, $c_m$, $\forall m\in\{1,2\}$, must be positive, otherwise $M_{1,(m)}$ will be zero. 

\subsubsection{Optimization Without Latency Constraints}
The first optimization problem can then be formulated as follows:
\begin{IEEEeqnarray*}{rClC}
	{\bf P1:} &\underset{\beta_1,\beta_2}{\text{max}} & \quad M_{1,(2)}                              \label{C1_OPT1}   \\
	&\text{subject to}              & \quad  {\bf C1:}  \quad M_{1,(1)}>M_{1,(1)}^{\text{target}},   \label{C2_OPT1}   \\ 
	&                               &  \quad {\bf C2:}  \quad   0<\beta_1<\frac{1}{2},   \qquad{\bf C3:}  \quad \beta_1+\beta_2=1.              \label{C3_OPT1}
\end{IEEEeqnarray*}
{\bf C2} ensures that the user with poor channel can decode its signal without any SIC  $0<\beta_1<\beta_2<1$ and {\bf C3} denotes the maximum BS power constraint $\beta_1+\beta_2=1$. 
Note that  $M_{1,(1)}$ and $M_{1,(2)}$ are decreasing functions of $A_{1,1}$ and $A_{1,2}$, respectively. $M_{1,(1)}>M_{1,(1)}^{\text{target}}$ imposes an upper bound on $A_{1,1}$ and maximizing $M_{1,(2)}$ is equivalent to minimizing $A_{1,2}$. Moreover, since $A_{1,1}$ and $A_{1,2}$ are increasing functions of $c_1$ and $c_2$\footnote{Note that $c_1$ and $c_2$ must be positive; otherwise, the first moments will be zero according to {\bf Theorem \ref{Thm2}}.}, we can  transform {\bf P1} as follows:
\begin{IEEEeqnarray*}{rClC}
{\bf P1:} & \underset{\beta_1,\beta_2}{\text{min}} &  c_2                             \\
&\text{subject to}     &\quad         0<c_1<c_1^{\text{target}},                 \\   && 0<\beta_1<\frac{1}{2},   \quad c_2>0,   \quad   \beta_1+\beta_2=1,~\nonumber                        
\end{IEEEeqnarray*}
where $0<c_2$ and $0<\beta_1$ guarantee positive $M_{1,(2)}$ and $M_{1,(1)}$, respectively, and $c_1^{\text{target}}$ can be obtained by solving the following equality:
\begin{IEEEeqnarray}{rCl}
    c_1^{\text{target}} \frac{\delta}{1-\delta} \, _2F_1(1,1-\delta;2-\delta;-c_1^{\text{target}}) = 
    2\left( \frac{1}{ M_{1,(1)}^{\text{target}} } -1 \right).
    \label{eq:c_target}
\end{IEEEeqnarray} 
Note that $c_1^{\text{target}}$ is unique and positive and can be obtained numerically. Moreover, since $c_1= \theta/\beta_1 $ and 
$c_2=\theta/(1-\beta_1(1+\theta))$, $c_2>0$ can be written as $\beta_1<1/(1+\theta)$ and $0<c_1<c_1^{\text{target}}$ can be written as
 $\beta_1>\theta/c_1^{\text{target}}$. {\bf P1} can then be rewritten as follows:
\begin{IEEEeqnarray}{rClC}
	{\bf P1:} & \underset{\beta_1}{\text{min}} & \beta_1                                                         \\
	&\text{subject to}     &\quad          
	\frac{\theta}{ c_1^{\text{target}} } < \beta_1 <  \min \left\{ \frac{1}{2},\frac{1}{1+\theta} \right\}.  \nonumber                
\end{IEEEeqnarray}
The aforementioned optimization problem can be solved in closed-form as follows.
\begin{Corollary}
When the problem is feasible, i.e., ${\theta}< c_1^{\text{target}} \min \left\{ \frac{1}{2},\frac{1}{1+\theta} \right\}$, 
the optimal powers for users can be obtained as $\beta_1^*=\theta / c_1^{\text{target}} $ and $\beta_2^*=1-\beta_1^*$, where $c_1^{\text{target}}$ is given in \eqref{eq:c_target}.
\end{Corollary}

\subsubsection{Optimization with Latency Constraints}
In URLLC systems, the local delay of a user is a crucial performance metric; therefore, in the following, we also consider the mean local delay constraints for each user.
\begin{IEEEeqnarray*}{rClC}
	{\bf P2:} &\underset{\beta_1,\beta_2}{\text{max}} & M_{1,(2)}                              \nonumber         \\
	&\text{subject to}              & \quad {\bf C1:} \quad M_{1,(1)}>M_{1,(1)}^{\text{target}},   \nonumber         \\ 
	&                               & \quad {\bf C2:} \quad   c_1 \frac{\delta}{1-\delta} < 2,  \quad      	    c_2 \frac{\delta}{1-\delta} < 1,       \label{C3_OPT2}   \\
	&                                    & \quad {\bf C3:} \quad  0<\beta_1<\frac{1}{2}, \quad\beta_1+\beta_2=1.                 \nonumber
\end{IEEEeqnarray*}
The constraints in {\bf C2} are the constraints for finite mean local delays for downlink NOMA users. Using  the constraints in {\bf C2} and {\bf C3}, {similar to {\bf P1}}, we can transform {\bf P2} as follows:
\begin{IEEEeqnarray*}{rClC}
		{\bf P2:} &\underset{\beta_1, \beta_2}{\text{min}} & c_2                         
	\nonumber \\
	&\text{subject to}             & \quad {\bf C1:} \quad  0<c_1< \min\left\{ 2\frac{1-\delta}{\delta},c_1^{\text{target}} \right\}, 
	\nonumber \\
	&                               & \quad {\bf C2:} \quad 0<c_2<\frac{1-\delta}{\delta},      
 \\
	&                                    & \quad {\bf C3:} \quad  0<\beta_1<\frac{1}{2}, \quad\beta_1+\beta_2=1,                 \nonumber                        
\end{IEEEeqnarray*}
and, finally, we can rewrite {\bf P2} as follows: 
\begin{IEEEeqnarray}{rClC}
	{\bf P2:}&\underset{\beta_1}{\text{min}} & \beta_1                                                         \nonumber \\
	&\text{subject to}              & \quad
	\frac{\theta}{ \min\left\{ 2\frac{1-\delta}{\delta},c_1^{\text{target}} \right\} } 
	< \beta_1 <
	\min \left\{ \frac{1}{2},\frac{ 1-\theta {\delta}/{(1-\delta)} }{1+\theta} \right\}.  \nonumber                
\end{IEEEeqnarray}
The aforementioned optimization problem can be solved in closed-form as follows.
\begin{Corollary}
When the problem is feasible, the optimal powers  are 
$\beta_1^*=\theta / \min\left\{ 2\frac{1-\delta}{\delta},c_1^{\text{target}} \right\} $ and $\beta_2^*=1-\beta_1^*$, where $c_1^{\text{target}}$ is obtained by \eqref{eq:c_target}. 
\end{Corollary}
Specifically,  considering the constraints of finite mean local delays decreases the feasible regions and changes the optimal power solutions.

\subsection{Transmit Power Optimization in $N$-User NOMA}
{We extend {\bf P2} for an $N$-user downlink NOMA network as follows:
\begin{IEEEeqnarray*}{rClClC}
{\bf P3:}	&\underset{\beta_1,\beta_2,\cdots,\beta_N} {\text{max}}  &\qquad& M_{1,(m)}                 &    &                        
	\nonumber \\
	&\text{subject to}              & \quad {\bf C1:}    & M_{1,(k)}>M_{1,(k)}^{\text{target}},           &    & k=1,2,\cdots,N,
	\nonumber \\
	&                               &  \quad {\bf C2:} \quad  & D_{1,k}=c_k \frac{\delta}{1-\delta} < N-k+1,   &    & k=1,2,\cdots,N,   \label{eq:c3_generalized} \\ 
	&                               &  \quad {\bf C3:}   & c_k=\left( \frac{\beta_k}{\theta} - \sum_{i=1}^{k-1} \beta_i \right)^{-1}>0   &    & k=1,2,\cdots,N,   
	\label{eq:c4_generalized} \\              
	&                               &\quad  {\bf C4:}   &  \beta_i \le \beta_j                           &    & \forall i,j \in \{ 1,\cdots,N \},
	                                                                                              \quad i \le j,     
	\label{eq:c5_generalized} \\              
	&                               &  \quad {\bf C5:}  &   \sum_{i=1}^{N} \beta_i =1,    \quad 0 \le \beta_k   &&                            k=1,2,\cdots,N,        \label{eq:c7_generalized}                                  
\end{IEEEeqnarray*}
where {\bf C1} denotes the minimum success probability constraint for each user\footnote{When there is no minimum success probability constraint for users at rank $m$, we can set $M_{1,(m)}^{\rm target}=0$.}, {\bf C2} represents the finite mean local delay constraints for all users, {\bf C3} guarantees positive success probability $M_{1,(k)}$ for each user, {\bf C4} and {\bf C5 } are  power constraints of the downlink NOMA system. According to {\bf Theorem \ref{Thm1}}, $M_{1,(k)}=\frac{N!}{(N-k)!} \prod_{i=1}^{k}\frac{1}{A_{1,k}+N-i+1}$, where $A_{1,k}=c_k \frac{\delta}{1-\delta}\,_2F_1(1,1-\delta;2-\delta;-c_k)$. 
$M_{1,(k)}$ is a decreasing function of $A_{1,k}$, and $A_{1,k}$ is an increasing function of $c_k$ where 
$c_k=\left( {\beta_k}/{\theta} - \sum_{i=1}^{k-1} \beta_i \right)^{-1}$.
Therefore, $M_{1,(k)}$ is a decreasing function of $c_k$ and maximizing $M_{1,(k)}$ is equivalent to maximizing $c_k^{-1}$. Moreover, $M_{1,(k)}>M_{1,(k)}^{\text{target}}$ can also be written as $c_k<c_k^{\text{target}}$, where $c_k^{\text{target}}$ is obtained by solving the following equation: 
\begin{IEEEeqnarray}{rCl}
	M_{1,(k)}^{\text{target}}=\frac{N!}{(N-k)!} \prod_{i=1}^{k}
	\frac{1}{c_k^{\text{target}} \frac{\delta}{1-\delta}\,_2F_1(1,1-\delta;2-\delta;-c_k^{\text{target}})+N-i+1}.
	\label{eq:c_target_generalized}
\end{IEEEeqnarray}
The above equation has a positive unique solution which can be obtained numerically. Combining $c_k<c_k^{\text{target}}$ and {\bf C2} yields $c_k<\min \left\{  \frac{1-\delta}{\delta}(N-k+1),c_k^{\text{target}} \right\}$ and {\bf P3} can be reformulated as:
{\begin{align}
	{\bf P3:} &\quad \underset{\beta_1,\beta_2,\cdots,\beta_N} {\text{max}}  \quad \frac{\beta_m}{\theta} - \sum_{i=1}^{m-1} \beta_i                                                                                                                      
	\nonumber \\
	& \text{subject to}  \quad 
{1}/{\min \left\{  \frac{1-\delta}{\delta}(N-k+1),c_k^{\text{target}} \right\}}<\frac{\beta_k}{\theta} - \sum_{i=1}^{k-1} \beta_i        \quad \forall k=1,2,\cdots,N,
	\nonumber \\        
	&      
\qquad \qquad \sum_{i=1}^{N} \beta_i =1, \quad \beta_1 \geq 0, \quad 0 \le \beta_{k}-\beta_{k-1}, \quad            \forall k=2,\cdots, N.     
	\nonumber                                  
\end{align}}
The optimal power allocation for {\bf P3}  can be obtained by using the linear programming techniques. Note that the formulated optimization problems {\bf P1, P2,} and {\bf P3} and their respective solution approaches are general to optimize the success probability of any user at $m$-th rank.
}

\section{Numerical Results and Discussions}
In this section, we present numerical and simulation results to validate the accuracy of the derived expressions. We also analyze the optimal power solutions obtained from {\bf P1, P2,} and {\bf P3}. Specifically, for uplink NOMA, we validate and compare the analytical results of {\bf Theorem~\ref{Thm1}} considering the two proposed  models for the interferers' point process. A comparison is also  provided with the traditional OMA scheme. For both uplink and downlink NOMA, we validate the accuracy of the beta approximation for the meta distribution using the results in {\bf Theorem \ref{Thm1}} and {\bf Theorem \ref{Thm2}} and show the distribution of the CSP for different users in a NOMA cluster. Finally, we  show the  impact of including user latency constraints in downlink transmission success  probability maximization problems. The optimal power solutions are illustrated  for various scenarios.

\subsection{Uplink NOMA}
\subsubsection{Validation of Model 1 and Model 2 and Meta Distribution of CSP} 
\begin{figure}[ht]
	\parbox[c]{.5\textwidth}{%
		\centerline{\subfigure[First moment of the CSP.]
			{\epsfig{file=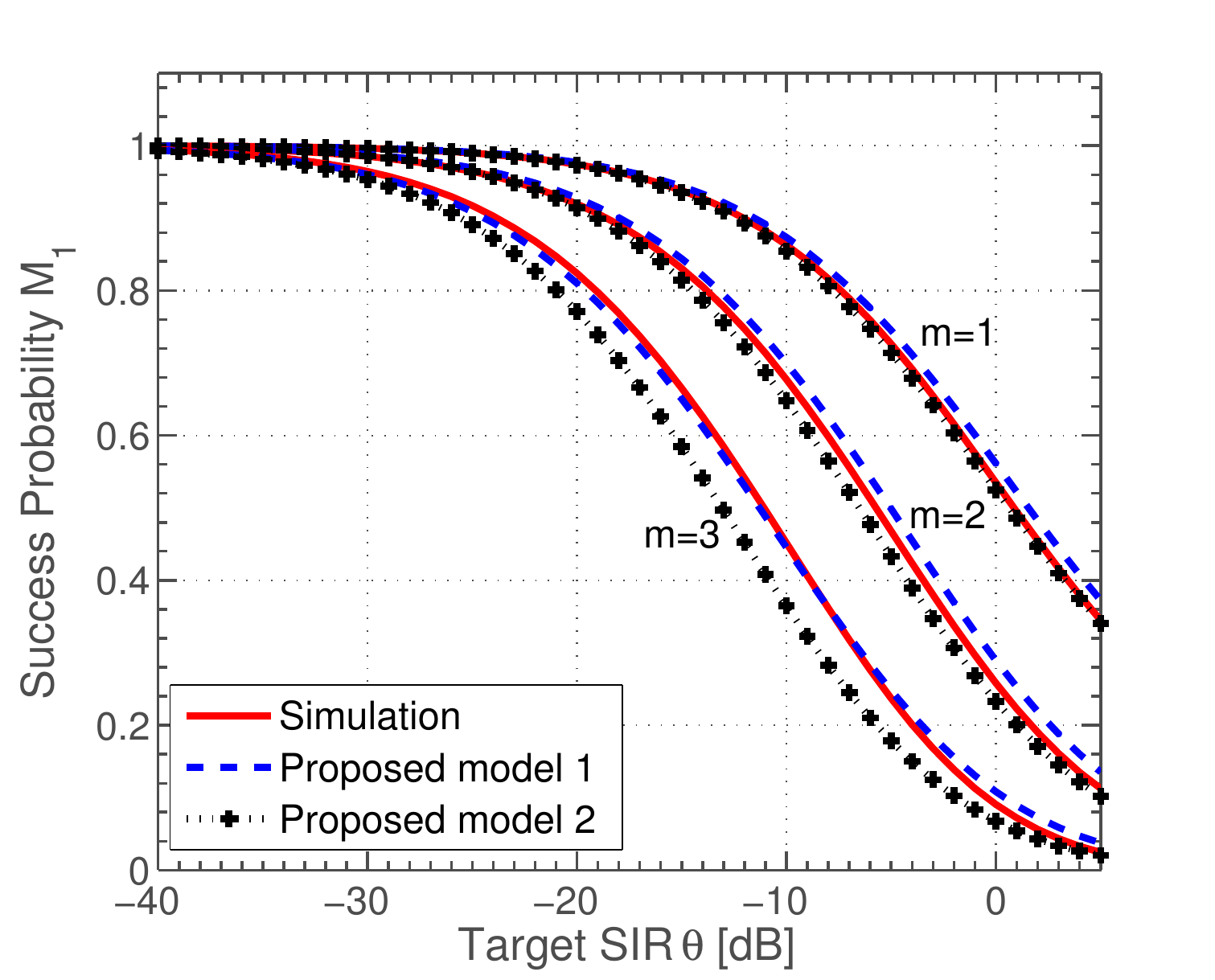, width=.48\textwidth}}}} 
\parbox[c]{.5\textwidth}{%
		\centerline{\subfigure[Meta distribution for $\theta=-5\,{\rm dB}$.]
			{\epsfig{file=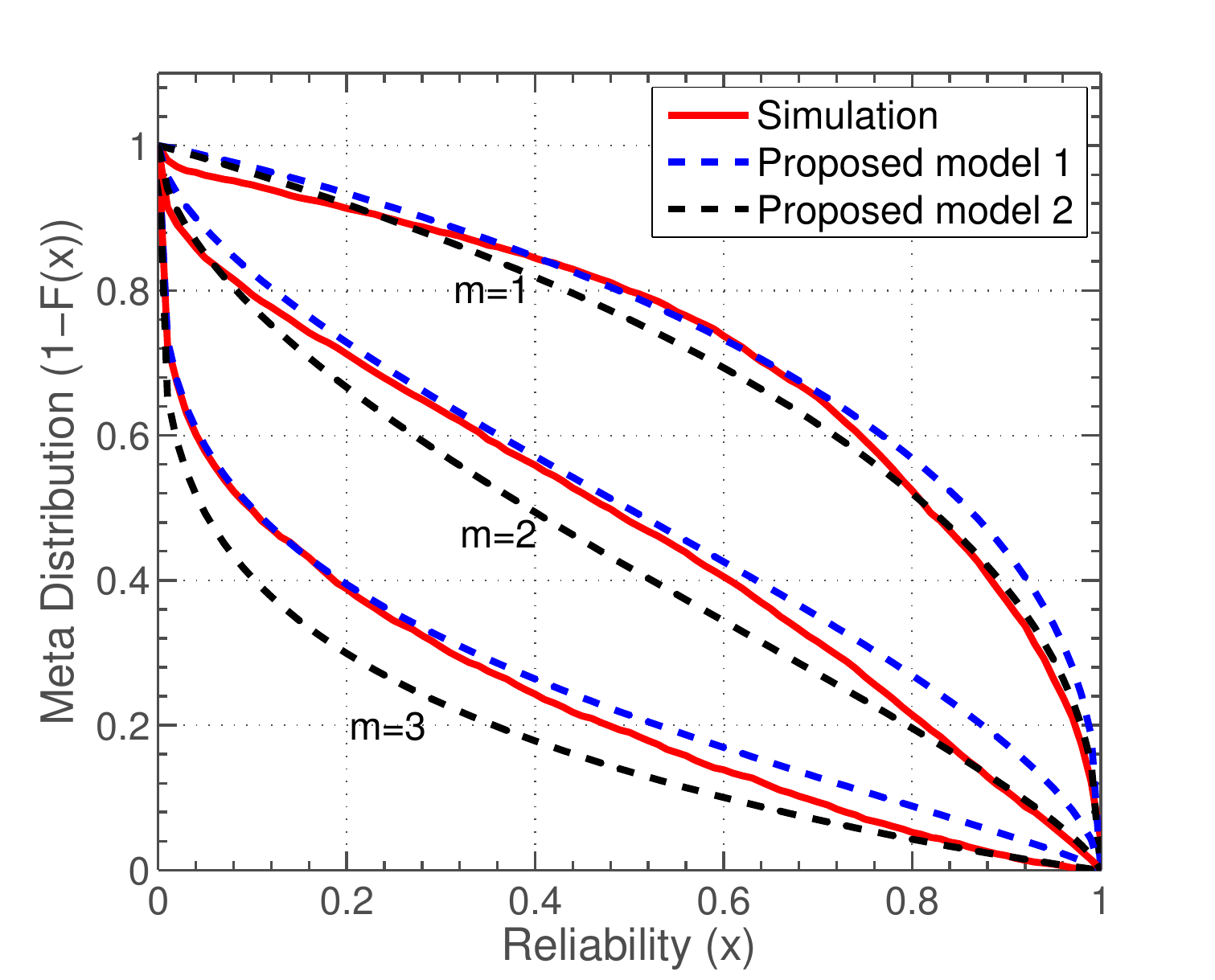, width=.48\textwidth}}}}
	\caption{First moment of the CSP and its meta distribution for three-users uplink NOMA for $\lambda_{\rm b}=0.001$, and $\alpha=4$.}
	\label{fig:success_probability_uplink}
\end{figure}

To demonstrate the accuracy of the proposed interferers' point process models, in \figref{fig:success_probability_uplink}(a), we plot the first moment of the CSP, which is the standard success probability, of a user at rank $m$. Simulation results and the analytical results derived in {\bf Theorem \ref{Thm1}} are compared for $\lambda_{\rm b}=0.001$, $N=3$, and $\alpha=4$. According to \figref{fig:success_probability_uplink},  \textbf{Model 2} provides a better approximation for $m=1$ while \textbf{Model 1} provides a better approximation for $m=N$. In general, \textbf{Model~1} outperforms in a wide range of scenarios. Also,  the closest user has the highest success probability compared to any other user in the typical Voronoi cell. 

{For the same network parameters, the exact meta distribution of the CSP (obtained via simulations) and its beta approximation (with two approximate interferers' point processes) are shown in \figref{fig:success_probability_uplink}(b). Using the proposed point process models, beta distribution provides a good approximation for the exact meta distribution; therefore, our expressions can be used to study the distribution of the CSPs in uplink NOMA.}


\subsubsection{NOMA vs. OMA} 
\begin{figure}
	\parbox[c]{.5\textwidth}{%
		\centerline{\subfigure[Uplink.]
			{\epsfig{file=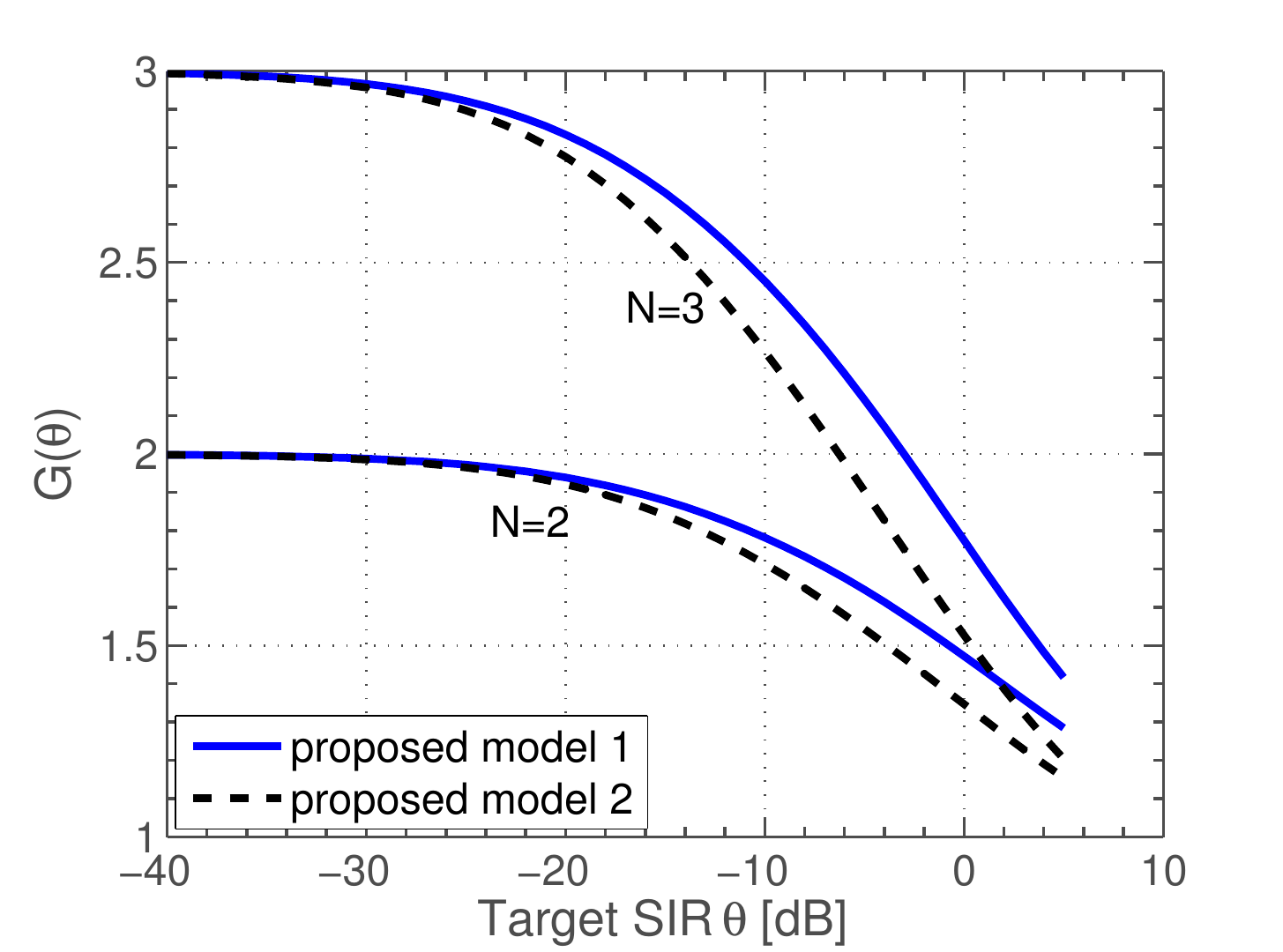,width=.48\textwidth}}}} 
	\parbox[c]{.5\textwidth}{%
		\centerline{\subfigure[Downlink.]
			{\epsfig{file=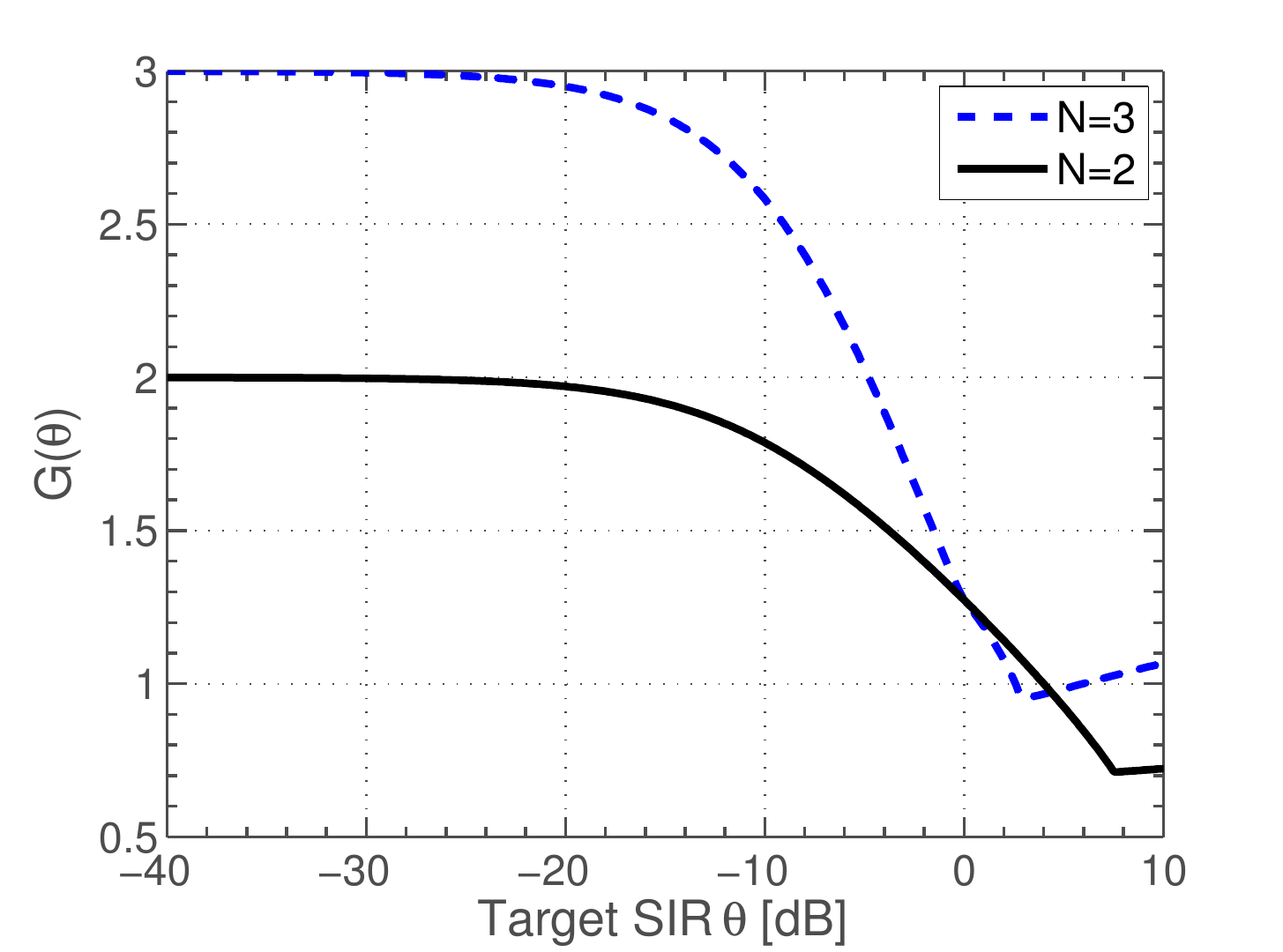,width=.48\textwidth}}}} 
	\caption{ $G(\theta)$ for uplink and downlink NOMA. (a) Uplink NOMA with $\lambda_{\rm b}=0.0005$ and $\alpha=4$.
		                  (b) Downlink NOMA with $\lambda_{\rm b}=0.001$ and $\alpha=4$. 
		                  For $N=2$, $\beta_1=0.15$ and $\beta_2=0.85$. 
		                  For $N=3$, $\beta_1=0.17$, $\beta_2=0.33$, and $\beta_3=0.5$.}
	\label{fig:gain}
\end{figure}

To compare $N$-user NOMA with OMA, we define the gain $G$ as
\begin{IEEEeqnarray}{rCl}
	G(\theta) \triangleq \frac{ \sum_{m=1}^{N} M_{1,(m)}(\theta) } { M_1^{\text{OMA}}(\theta) }, \label{eq:gain}
\end{IEEEeqnarray} 
where $M_1^{\text{OMA}}$ considers  no channel inversion power control and is obtained by setting $N=m=1$ in {\bf Theorem \ref{Thm1}}. For a given amount of radio bandwidth, {when the user point process is ergodic, $G(\theta)$ can be interpreted as the ratio of the density of users served in NOMA to the density of users served in OMA.} 
For instance, according to \figref{fig:gain}(a), when $N=3$, $G(-10~{\rm dB}) \approx 2.3$, which means, with NOMA, the number of users served in a unit area is 2.3 times that with OMA. In \figref{fig:gain}(a),  the gain of uplink NOMA $G(\theta)$ decays rapidly with increasing $\theta$ and the rate of decay is much higher for large number of users $N$. 

\subsection{Downlink NOMA}
\subsubsection{NOMA vs OMA} In \figref{fig:gain}(b),  $G(\theta)$ is evaluated for downlink. 
Similar to the uplink, the gain of downlink NOMA $G(\theta)$ decays rapidly with increasing $\theta$ and the rate of decay is much higher for large number of users $N$. However, for large values of $\theta$, $G(\theta)$ increases since the effect of link distance is dominant and the average link distance of a typical user in OMA is $1/(2\sqrt{\lambda_{\rm b}})$, while in NOMA, the average link distance of the 1-st rank user is $1/(2\sqrt{N\lambda_{\rm b}})$.

\subsubsection{Validation of Meta Distribution of CSP} 
\begin{figure}
	\parbox[c]{.5\textwidth}{%
		\centerline{\subfigure[$\beta_1=0.17$, $\beta_2=0.33$, and $\beta_3=0.5$.]
			{\epsfig{file=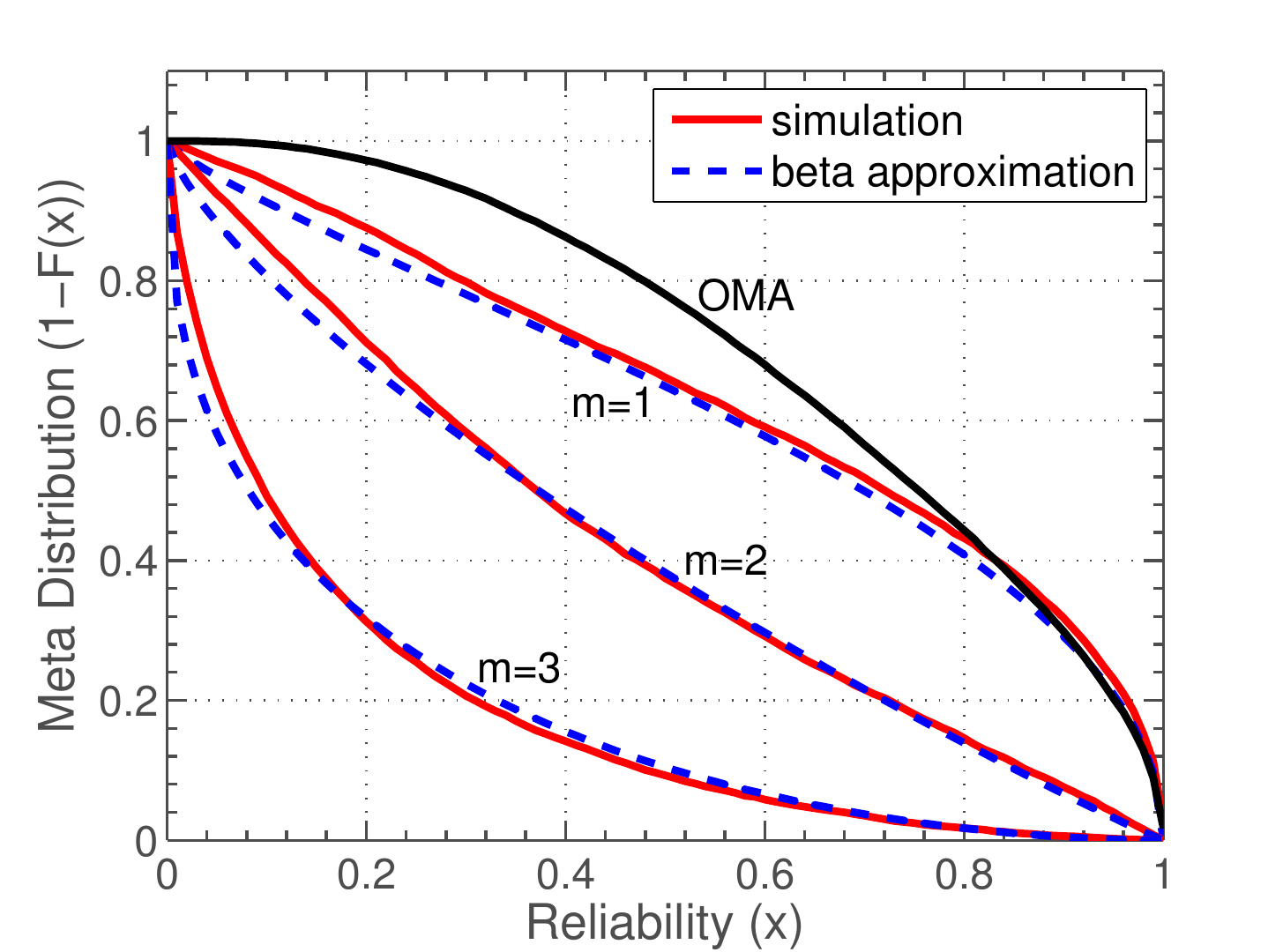,width=.48\textwidth}}}} 
	\parbox[c]{.5\textwidth}{%
		\centerline{\subfigure[$\beta_1=0.08$, $\beta_2=0.25$, and $\beta_3=0.67$.]
			{\epsfig{file=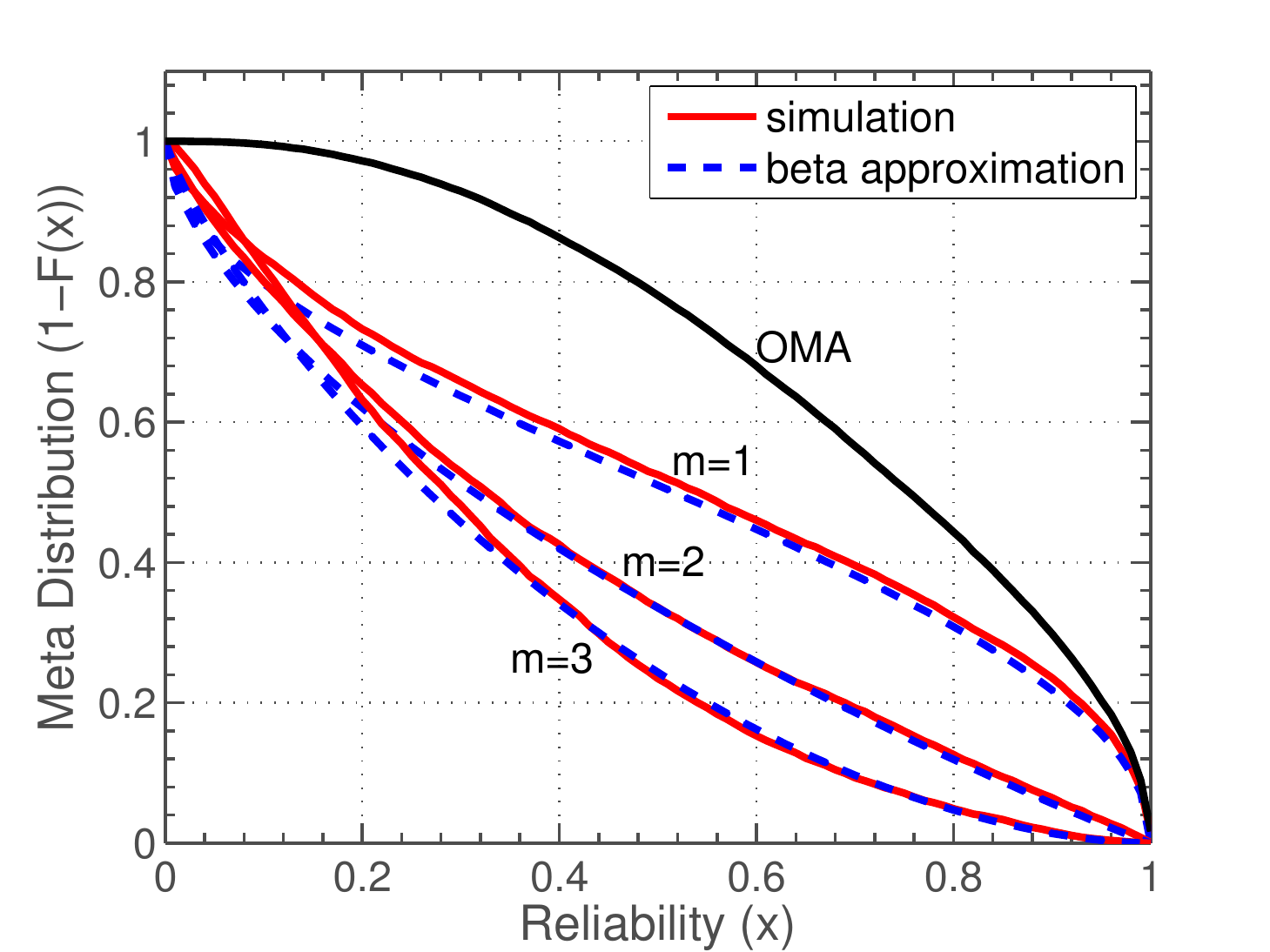,width=.48\textwidth}}}} 
	\caption{Beta approximation and the exact meta distribution for $m$-th rank user when $\lambda=0.001$, $N=3$, $\alpha=4$, and $\theta=-3\,{\rm dB}$.}
	\label{fig:beta_accuracy}
\end{figure}

In \figref{fig:beta_accuracy}, we show that the meta distribution for the CSP can be approximated by the beta distribution with shape parameters ${M_1 \beta}/{(1-M_1)}$ and $\beta$. We consider three users in each NOMA cell. In this scenario, the meta distribution of the $m$-th rank user, $m=1,2,3$, and its beta approximation are shown in \figref{fig:beta_accuracy} for two different power allocations. It can be seen that the beta distribution provides a good approximation for the meta distribution.
{In \figref{fig:beta_accuracy}(a),} we note that about $58\%$ of the 1-st rank users, $30\%$  of 2-nd rank users, and $7\%$ of 3-rd rank users have success probabilities greater than 0.6. Therefore, success probabilities of $(58+30+7)/3\approx32\%$ of users are greater than 0.6. With OMA, for $68\%$ of users, success probabilities are greater than 0.6. This means that, with NOMA, the density of users served with the same amount of radio spectrum is $32\times 3/68 \approx 1.4$ times  that with OMA, when the success probability is higher than 0.6 (i.e., with reliability 0.6).

Moreover, using \eqref{distributionLD}, we can also study the distribution of the local delay from \figref{fig:beta_accuracy}. We note that $58\%$ of the $1$-st rank users successfully receive their desired signals with probability more than $0.6$ in the first time slot, while, after the second time slot, $74\%$ of the $1$-st rank users successfully receive their desired signals with probability more than $0.6$ (this is obtained by setting $k=2$ and $x=0.6$ in \eqref{distributionLD} which yields $\bar{F}_{P_{{\rm s},m}}(0.37)$). This value for the $2$-nd rank users is $50\%$ and for the $3$-rd rank users is $17\%$. Hence, after the second time slot, $(74+50+17)/3\approx47\%$ of users receive their desired signals with reliability $0.6$.

\subsubsection{{Finite and Infinite Mean Local Delay}} 
\begin{figure}
	\parbox[c]{.5\textwidth}{%
		\centerline{\subfigure[$\beta_1=0.35$ and $\beta_2=0.65$.]
			{\epsfig{file=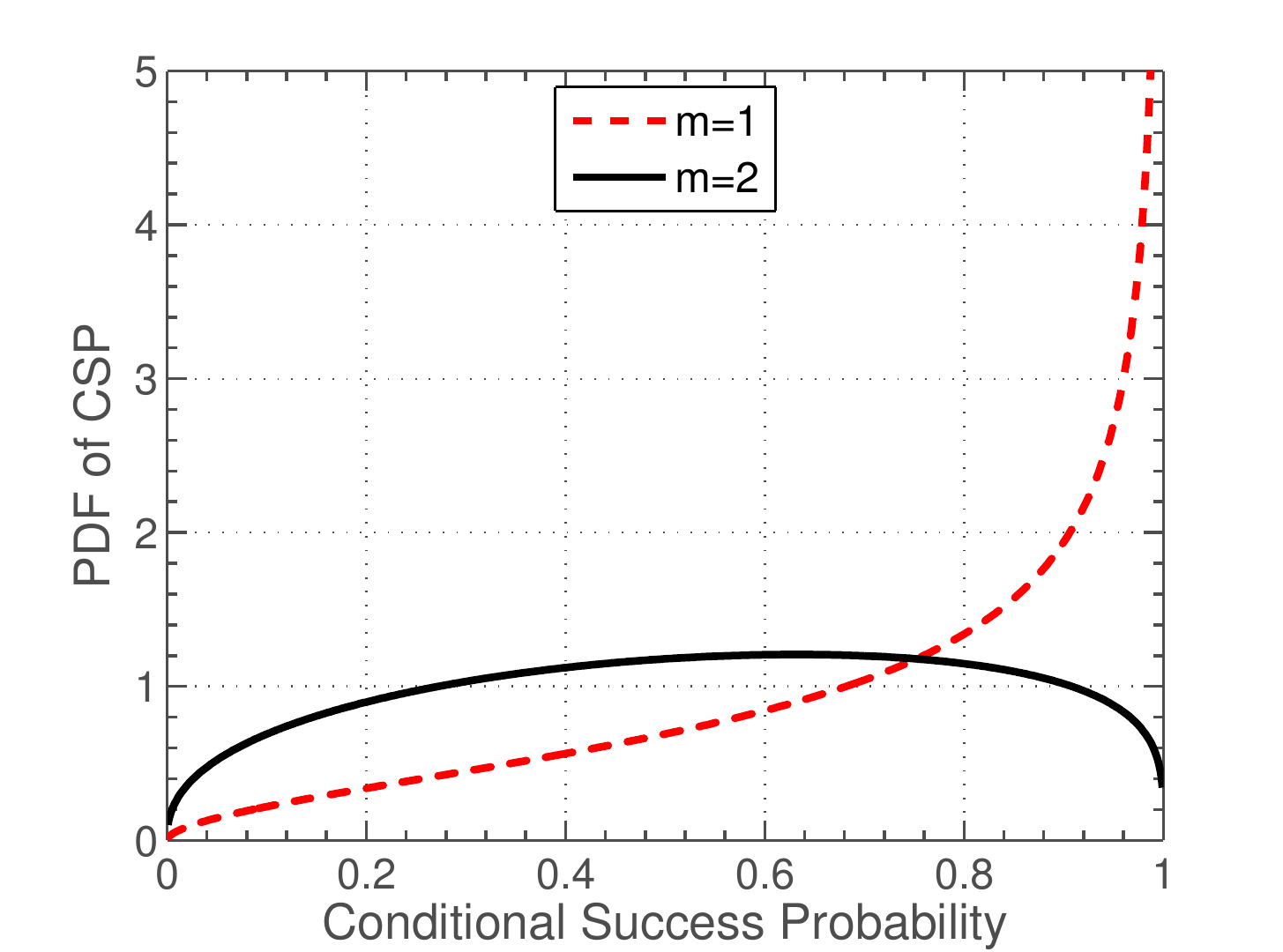,width=.48\textwidth}}}} 
	\parbox[c]{.5\textwidth}{%
		\centerline{\subfigure[$\beta_1=0.15$ and $\beta_2=0.85$.]
			{\epsfig{file=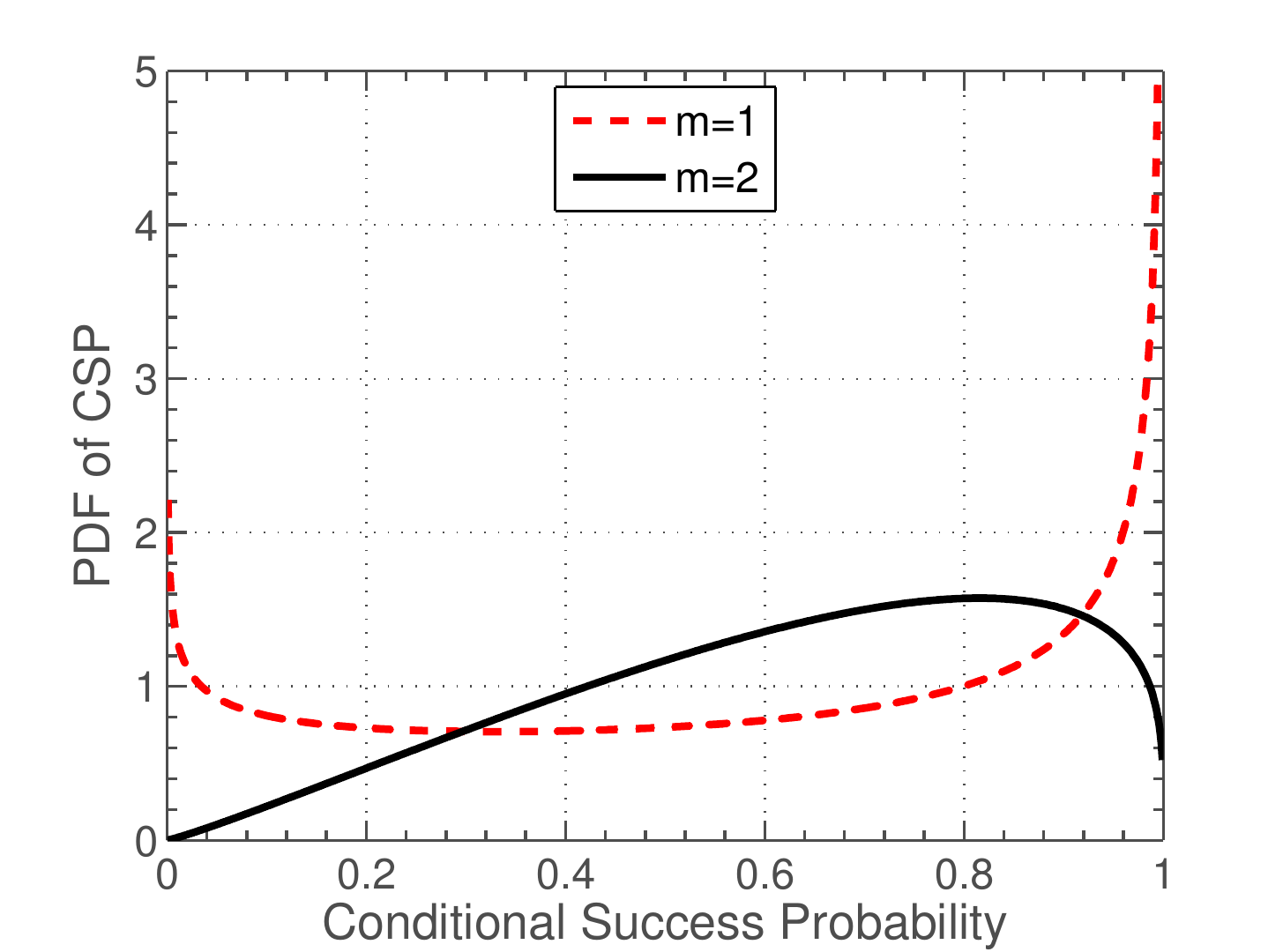,width=.48\textwidth}}}} 
	\caption{PDF of the CSP of $1$-st and $2$-nd rank downlink NOMA users for $\lambda_{\rm b}=0.001$, $N=2$, $\alpha=4$, $\theta=-5\,{\rm dB}$. (a) $M_{1,(1)}=0.73$ and $M_{1,(2)}=0.53$. (b) $M_{1,(1)}=0.59$ and $M_{1,(2)}=0.63$.}
	\label{fig:finite_and_infinite_mean_local_delay}
\end{figure}

Using the beta approximation, the distribution of the CSP of $1$-st and $2$-nd rank users are shown in \figref{fig:finite_and_infinite_mean_local_delay}.  To understand the relations between the CSP of users, the standard success probability ($1$-st moment), and the mean local delay ($-1$-st moment) consider the following examples. 

When $\lambda_{\rm b}=0.001$, $N=2$, $\alpha=4$, $\theta=-5\,{\rm dB}$, $\beta_1=0.35$,  and $\beta_2=1-0.35=0.65$, the standard success probability for the $1$-st rank users is 0.73 and for the $2$-nd rank users is 0.53. For the $1$-st and $2$-nd rank users, the mean local delays are finite, i.e., $c_m \frac{\delta}{1-\delta}<N-m+1$ is satisfied for $m=1$ and $m=2$.
When $\beta_1=0.15$ and $\beta_2=0.85$, the standard success probability for the $1$-st rank users is 0.59 and for the $2$-nd rank users is 0.63. Although the standard success probabilities are close, for the $1$-st rank users the mean local delay is infinite while for the $2$-nd rank users the mean local delay is finite. When the mean local delay is infinite, it means that there is a significant number of users with small conditional success probabilities in the network \cite{Baccelli2010}. This can also be seen in \figref{fig:finite_and_infinite_mean_local_delay}(b) where the PDF of small values of CSP for the $1$-st rank users is not zero. Therefore, we can conclude that for the $1$-st rank users CSPs are close to 0 and 1 with high probability while for the $2$-nd rank users they are close to mean 0.63 with high probability.

\subsubsection{Optimal Power Solutions} 
\begin{figure}[ht]
	\parbox[c]{.5\textwidth}{%
		\centerline{\subfigure[Optimal power of the first rank users $\beta_1^*$ as a function of target SIR $\theta$.]
			{\epsfig{file=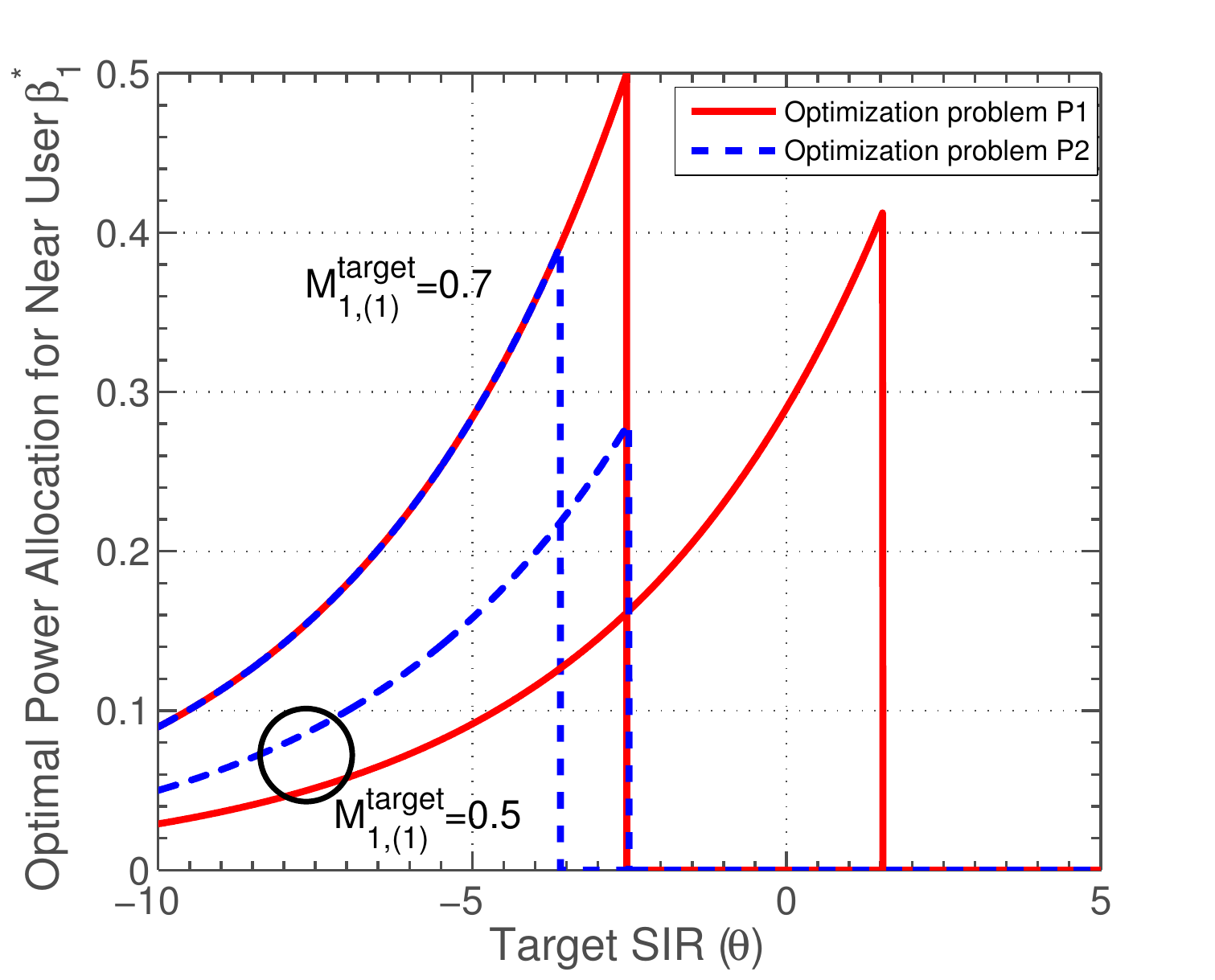,width=.48\textwidth}}}} 
	\parbox[c]{.5\textwidth}{%
		\centerline{\subfigure[Mean success probability of the second rank users as a function of target SIR $\theta$.]
			{\epsfig{file=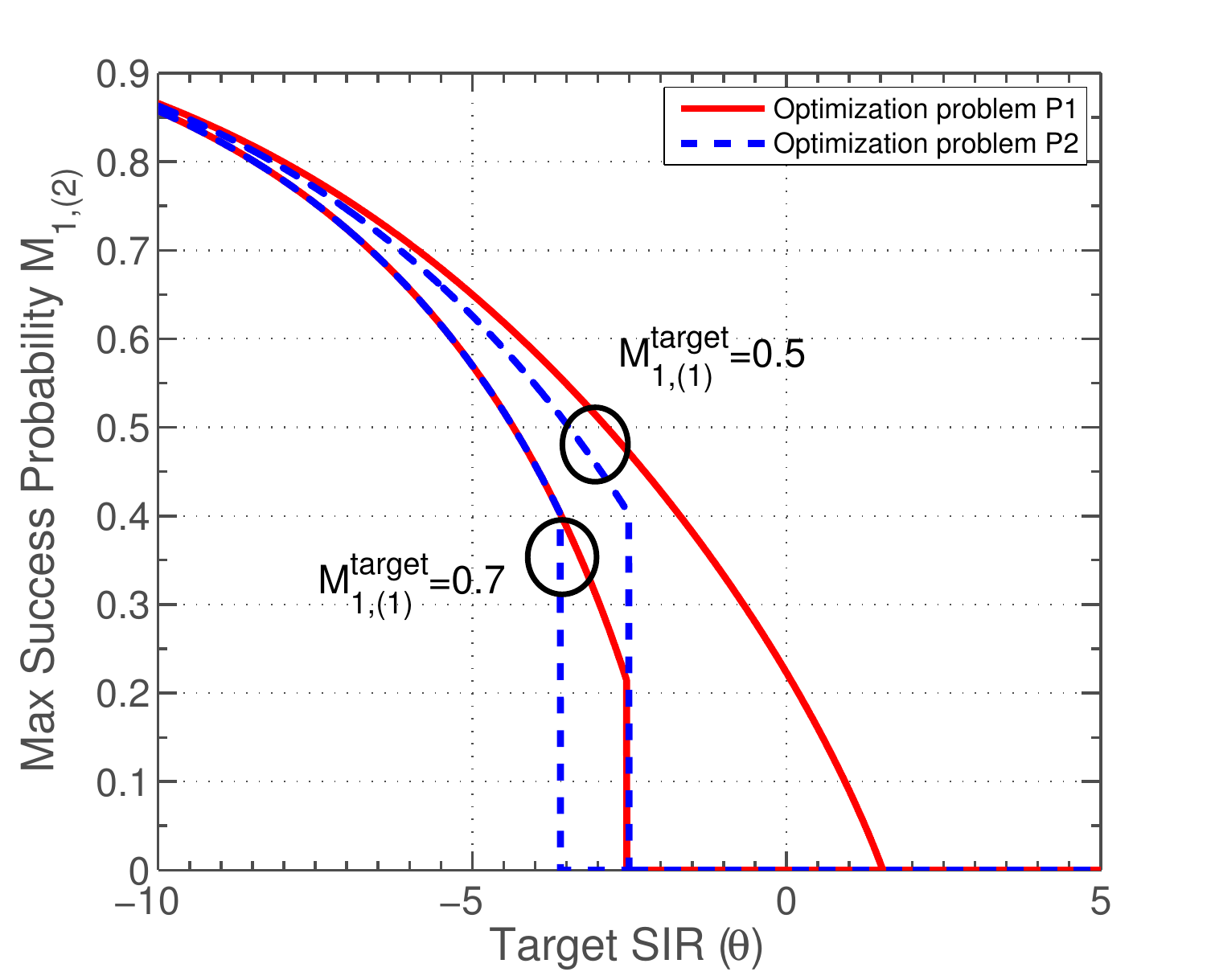,width=.48\textwidth}}}} 
	\caption{Comparison between the first and second optimization problems for $\alpha=4$. Zero values of $\beta_1^*=0$ 
		     (or $\text{max}\,M_{1,(2)}$) mean that the optimization problem is infeasible.}
	\label{fig:optimization}
\end{figure}

\figref{fig:optimization} shows the optimal powers of users as well as the maximum success probability achieved at the  2-nd rank user ($M_{1,(2)}$) as a function of target SIR $\theta$ for the first optimization problem ({\bf P1}) and the second optimization problem ({\bf P2}). For $M_{1,(1)}^{\text{target}}=0.7$, when both problems are feasible (in \figref{fig:optimization}, zero values correspond to infeasible problems), the optimal powers are the same.
However, for $M_{1,(1)}^{\text{target}}=0.5$, when both problems are feasible, the optimal powers are different. For instance, when $\theta=-3\,\text{dB}$, for {\bf P1} we have $\beta_1^*\approx0.15$, $M_{1,(1)}\approx0.50$, and $M_{1,(2)}\approx0.51$ while {\bf P2} yields $\beta_1^*\approx0.25$, $M_{1,(1)}\approx0.60$, and $M_{1,(2)}\approx 0.46$. 

As we discussed in the previous section, although there is a small difference between the { achieved mean success probabilities for the 2-nd rank users (maximum $M_{1,(2)}$)}, as shown in \figref{fig:optimization}(b), there is a significant difference between the distributions of the CSP  and hence the optimal power solutions, as shown in \figref{fig:optimization}(a). Moreover, for the first optimization problem ({\bf P1}), a large number of the 1-st rank users have success probabilities close to 0 (and also close to 1). However, with the delay constraints in the second optimization problem ({\bf P2}), the success probabilities of the 1-st rank users become close to the mean value. 


\begin{figure}
	\parbox[c]{.5\textwidth}{%
		\centerline{\subfigure[Optimal powers.]
			{\epsfig{file=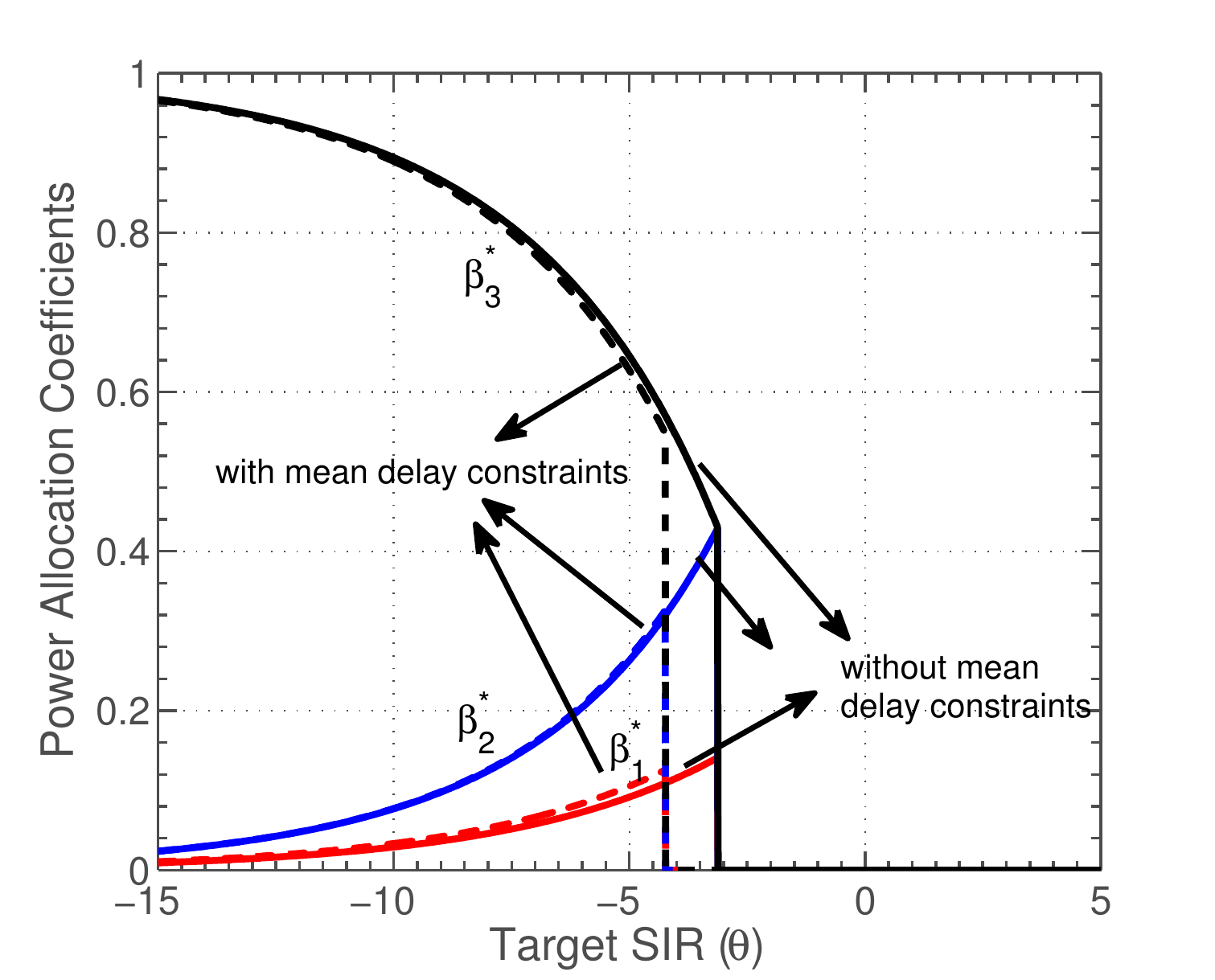,width=.48\textwidth}}}} 
	\parbox[c]{.5\textwidth}{%
		\centerline{\subfigure[Maximum $M_{1,(3)}$.]
			{\epsfig{file=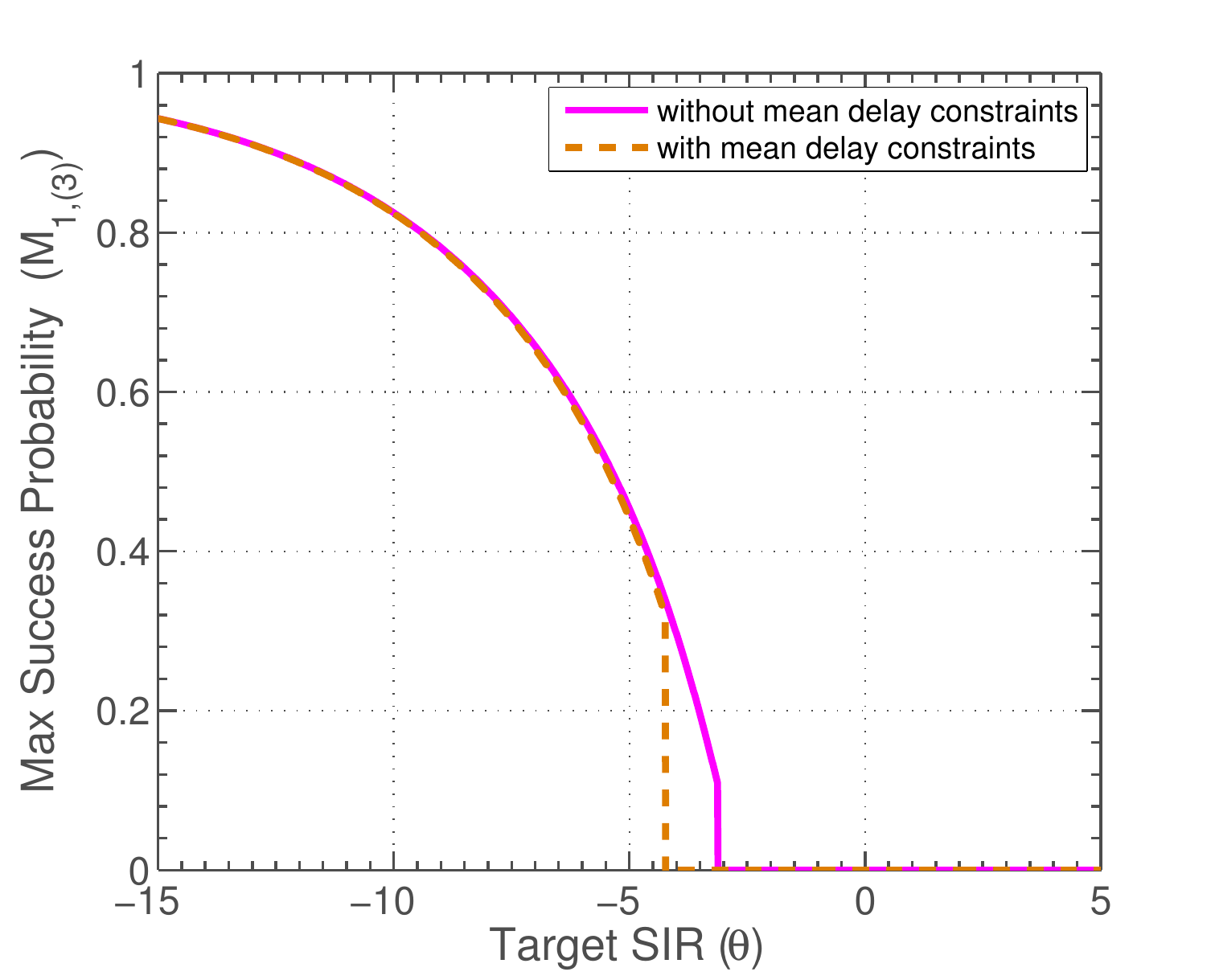,width=.48\textwidth}}}} 
	\caption{Optimal powers and maximum $M_{1,(3)}$ for 3-UE downlink NOMA with and without considering the finite mean local delay constraints for $M_{1,(1)}^{\text{target}}=0.6$, $M_{1,(2)}^{\text{target}}=0.5$, and $\alpha=4$.}
	\label{fig:optimization_2}
\end{figure}

For $N=3$, the optimal powers and maximum $M_{1,(3)}$ are illustrated in \figref{fig:optimization_2} when $M_{1,(1)}^{\text{target}}=0.6$ and $M_{1,(2)}^{\text{target}}=0.5$. The optimal powers and maximum $M_{1,(3)}$, when the finite mean local delay constraints are not considered, are also illustrated for comparison. When both problems are feasible, the maximum $M_{1,(3)}$  are the same for both the problems. However, considering the finite mean local delay constraints avoids small (zero and close to zero) CSPs for the 1-st rank users.

\section{Conclusion}
We have developed a stochastic geometry   framework  to derive the moments of the conditional success probability  (CSP) and its meta distribution in uplink and downlink NOMA networks. The CSP and its meta distribution are useful in the evaluation of the network performance metrics such as the standard success probability and average local delay. {For uplink NOMA}, we have proposed two point process models for the spatial locations of the interferers by using the definition of BS/user pair correlation function and demonstrated the accuracy of the models by using Monte-Carlo simulations. For downlink NOMA, we have derived closed-form solutions for the success probability, its meta distribution, and the average local delay. As an application of the developed analytical framework, we have used the closed-form results to optimize downlink transmit powers in order to maximize the success probability  with and without latency constraints. The optimal solutions have been obtained in closed-form for two-user downlink NOMA networks and these solutions reveal the significance of including the latency constraints in the traditional optimization problems. The framework can be extended for more advanced network models with Matern and Thomas cluster processes. Also, network performance can be optimized under constraints such as variance and kurtosis/skewness of the local delay and success probability. {Moreover, the impact of imperfect SIC in uplink and downlink NOMA can be studied. Analysis of the imperfect SIC is challenging since we need to consider the interference correlation.}

\section*{Appendix A: Proof of Theorem~1}
\renewcommand{\theequation}{A.\arabic{equation}}
\setcounter{equation}{0}
We first derive the CSP $P_{\rm s,(m)}$ for the $m$-th rank uplink NOMA user as follows:
		\begin{IEEEeqnarray}{rCl}
			P_{\rm s,(m)}(\theta) &=& \mathbb{P} \left( \SIR_{(m)} > \theta \mid \Phi_{\rm U},\text{tx} \right)	\nonumber
			\\
			&=& \mathbb{P} \left( h_{x_{(m)}} > \theta \|x_{(m)}\|^{\alpha} 
			\left( \sum_{x \in \Phi_{\rm I}} h_x \|x\|^{-\alpha} + \sum_{i=m+1}^{N} h_{x_{(i)}}\|x_{(i)}\|^{-\alpha} \right) \mid \Phi_{\rm U},\text{tx} \right) \nonumber\\
			&\stackrel{(a)}{=}& \prod_{x\in\Phi_{\rm I}} \frac{1}{ 1+\theta\|x_{(m)}\|^{\alpha}\|x\|^{-\alpha} }
			\prod_{i=m+1}^{N} \frac{1}{ 1+\theta\|x_{(m)}\|^{\alpha}\|x_{(i)}\|^{-\alpha} },
			\label{eq:CSP_uplink}
		\end{IEEEeqnarray}
where (a)  follows from applying the CCDF of the unit mean exponential distribution of $h_{x_{(m)}}$ and then the Laplace transform of the unit mean exponential distribution of $h_x$ and $h_{x_{(i)}}$. Note that $\Phi_{\rm U}$ represents the superposition of two independent point processes, namely, the inter-cell interferers' point process $\Phi_{\rm I}$ and the {point process of users located in the typical Voronoi cell (intra-cell users).}

Next, we derive the $b$-th moment of CSP $M_{b,(m)} = \mathbb{E}_{\Phi_{\rm U}} \left[ P_{\rm s,(m)}^b \right] $ as follows:
\begin{IEEEeqnarray}{rCl}
&M_{b,(m)} 
&\stackrel{(a)}{=}
\mathbb{E}_{x_{(m)}}
			\left[ 
			\underbrace{\mathbb{E}_{\Phi_{\rm I}} \left[ \prod_{x\in\Phi_{\rm I}} \left( \frac{1}{ 1+\theta\|x_{(m)}\|^{\alpha}\|x\|^{-\alpha} } \right)^b  \right]}_{{\bf Part \; A}}
			\underbrace{\mathbb{E}_{r} \left[ \left( \frac{1}{ 1+\theta\|x_{(m)}\|^{\alpha}r^{-\alpha} } \right)^b  \right]^{N-m}}_{{\bf Part \; B}}
			\right], \nonumber \\
			\label{eq:moment_CSP_uplink}
		\end{IEEEeqnarray}
where (a) is obtained by noting that (i)~$\Phi_{\rm U}$ is the superposition of the inter-cell and intra-cell point processes, {(ii)~the inter-cell interferers' point process and the intra-cell interferers' point process are independent, and (iii)}~conditioned on the user at rank $m$, the distribution of the distances of  the  intra-cell  interfering  users  from the typical BS  are i.i.d, so we can replace $||x_{(i)}||$ with $r$~\cite{Tabassum2017}.  Now using the {\bf Model 1} for inter-cell interferers point process in Section~IV, we can approximate {\bf Part A}  as follows:
		\begin{IEEEeqnarray}{rCl}
				\mathbb{E}_{\Phi_{\rm I}} \left[ \prod_{x\in\Phi_{\rm I}} \left( \frac{1}{ 1+\theta\|x_{(m)}\|^{\alpha}\|x\|^{-\alpha} } \right)^b  \right] \approx  \mathbb{E} \left[ \prod_{x\in\bar{\Phi}_{\rm P}} \left( \frac{1}{ 1+\theta\|x_{(m)}\|^{\alpha}\|x\|^{-\alpha} } \right)^{Nb} \mid x_{(m)} \right] 
\nonumber\\
\stackrel{(a)}{=} \exp \left(-2 \pi \lambda_{\rm b} \int\limits_0^{\infty} \left[1-\left( \frac{1}{ 1+\theta\|x_{(m)}\|^{\alpha}r^{-\alpha} } \right)^{Nb} \right] \left( 1- e^{-(12/5)\lambda_{\rm b}\pi r^2}\right) r {\rm d}r \right), 
		\end{IEEEeqnarray}
where we approximate $\Phi_{\rm I}$ with $\bar \Phi_{\rm I}$ and $\bar \Phi_{\rm I}$ is same as the parent process (which is PPP) with collocated $N$ daughters. The last equality is obtained from the probability generating functional (PGFL) of PPP. Similarly, using the {\bf proposed model 2}, where we approximate $\Phi_{\rm I}$ with $\dbar{\Phi}_{\rm I}$, {\bf Part A} can be derived as follows:
		\begin{IEEEeqnarray}{rCl}
				&\mathbb{E}_{\Phi_{\rm I}} \left[ \prod_{x\in\Phi_{\rm I}} \left( \frac{1}{ 1+\theta\|x_{(m)}\|^{\alpha}\|x\|^{-\alpha} } \right)^b \right] \approx
			\mathbb{E} \left[ \prod_{x\in\dbar{\Phi}_{\rm I}} \left( \frac{1}{ 1+\theta\|x_{(m)}\|^{\alpha}\|x\|^{-\alpha} } \right)^b \mid x_{(m)} \right] \nonumber
			\\&=
			\exp \left(-2 \pi N \lambda_{\rm b} \int\limits_0^{\infty} 
			\left[ 1-\left( \frac{1}{ 1+\theta\|x_{(m)}\|^{\alpha}r^{-\alpha} } \right)^{b} \right] 
			\left( 1- e^{-(12/5)\lambda_{\rm b}\pi r^2}\right) r {\rm d}r \right). 
		\end{IEEEeqnarray}
Now, {\bf Part B} in \eqref{eq:moment_CSP_uplink} is derived as follows:
		\begin{align}
			&\mathbb{E} \left[ \left( \frac{1}{ 1+\theta\|x_{(m)}\|^{\alpha}r^{-\alpha} } \right)^b \mid x_{(m)} \right] = 
			\int\limits_{\|x_{(m)}\|}^{\infty} \left( \frac{1}{ 1+\theta\|x_{(m)}\|^{\alpha}r^{-\alpha} } \right)^b 
			f_{R_{\rm out}}\left(r \mid \|x_{(m)}\|\right) {\rm d}r \nonumber
			\\
			&= (5/2) \lambda_{\rm b} \pi \|x_{(m)}\|^2 e^{(5/4) \lambda_{\rm b} \pi \|x_{(m)}\|^2} 
			\int_0^1 \frac{t^{-3} e^{- (5/4) \lambda_{\rm b} \pi \|x_{(m)}\|^2 t^{-2}}} {\left( 1+\theta t^{\alpha} \right)^b} {\rm d}t.
			\label{eq:intra_cell_interference_uplink}
		\end{align}
Finally, {\bf Theorem~1} is obtained by averaging over the desired link distance using \eqref{eq:PDFdistance_m_uplink}.

\section*{Appendix B: Proof of Theorem~2}
\renewcommand{\theequation}{A.\arabic{equation}}
\setcounter{equation}{0}
Substituting \eqref{eq:SIR_downlink} in \eqref{CSP}, the CSP for the $m$-th rank user yields
		\begin{IEEEeqnarray}{rCl}
			P_{{\rm s},(m)}(\theta) &=& \mathbb{P} \left( \frac{I_{(m)}^{\rm intra} + I_{(m)}^{\rm inter}}{ \beta_m P h_0 \|x_0\|^{-\alpha} } < \frac{1}{\theta} \mid \Phi_{\rm B},\text{tx} \right) \nonumber
			\\
			&\overset{ \text {(a)} }{\mathop{=}}& \mathbb{P} \left( \frac{I_{(m)}^{\rm inter}}{ \beta_m P h_0 \|x_0\|^{-\alpha} } < \frac{1}{\theta} - 
			\frac{\sum_{i=1}^{m-1} \beta_i}{\beta_m} \mid \Phi_{\rm B},\text{tx} \right) \nonumber
			\\
			&\overset{ \text {(b)} }{\mathop{=}}& \mathbb{E}_{h_x} \left[ \exp\left\{ -c_m \|x_0\|^{\alpha}
			\left( \sum_{x\in\Phi_{\rm B}\setminus\{x_0\}} h_x \|x\|^{-\alpha} \right) \right\} \right] \nonumber
			\\
			&=& \prod_{x\in\Phi_{\rm B}\setminus\{x_0\}} \frac{1}{1+c_m \|x_0\|^{\alpha}\|x\|^{-\alpha}}, \nonumber
		\end{IEEEeqnarray}
where (a) is obtained by using \eqref{eq:I_intra_downlink}. When 
${\beta_m}/{\theta} - {\sum_{i=1}^{m-1} \beta_i}$ is not positive, $P_{{\rm s},(m)}(\theta)=0$. Therefore, in the following, we consider ${\beta_m}/{\theta} - {\sum_{i=1}^{m-1} \beta_i} >0$. (b) follows from the exponential distribution of $h_0$, applying \eqref{eq:I_inter_downlink}, and setting $c_m = \left( \frac{\beta_m}{\theta}-\sum\limits_{i-1}^{m-1}\beta_i \right)^{-1}$. Using $P_{{\rm s},(m)}$, now we can derive $M_{b,(m)}$ as follows:
		\begin{IEEEeqnarray}{rCl}
			M_{b,(m)} &\overset{ \text {(a)} }{\mathop{=}}& 
			\mathbb{E} \left[ \prod_{x\in\Phi_{\rm B}\setminus\{x_0\}} \left( \frac{1}{1+c_m \|x_0\|^{\alpha}\|x\|^{-\alpha}} \right)^b \right] \nonumber
			\\
			&\overset{ \text {(b)} }{\mathop{=}}& \mathbb{E}_{R_m} \left[ \exp\left\{ -\int\limits_{\mathbb{R}^2\setminus b(o,R_m)} 
			\left[ 1-\left( \frac{1}{1+c_m R_{m}^{\alpha}\|x\|^{-\alpha}} \right)^b \right] \lambda_{\rm b} {\rm d}x \right\} \right] \nonumber
			\\
			&\overset{ \text {(c)} }{\mathop{=}}& \mathbb{E}_{R_m} \left[ \exp\left\{ -2\pi\lambda_{\rm b} \int\limits_{R_m}^{\infty} 
			\sum_{k=1}^{\infty} \binom bk (-1)^{k+1} c_m^k R_{m}^{\alpha k} \frac{ r^{-\alpha k+1} }{ (1+c_mR_m^{\alpha}r^{-\alpha})^k }
			{\rm d}r \right\} \right] \nonumber
			\\
			&\overset{ \text {(d)} }{\mathop{=}}& \int\limits_{0}^{\infty}  \exp\left\{ -\pi\lambda_{\rm b} r^2  
			\sum_{k=1}^{\infty} \binom bk (-1)^{k+1} c_m^k  \frac{\delta}{k-\delta} \, _2F_1(k,k-\delta;k-\delta+1;-c_m) \right\} f_{R_m}(r) {\rm d}r, \nonumber
		\end{IEEEeqnarray}
where the expectation in (a) is over the point process $\Phi_{\rm B}$, (b) follows from probability generating functional (PGFL) of PPP \cite{haenggi2012stochastic} outside $b(o,R_m)$, (c) is obtained by using the polar domain representation and by applying the binomial expansion, and finally, $M_{b,(m)}$ in \eqref{eq:moment_CSP_downlink} is obtained by calculating the integral in (d) where $f_{R_m}(r)$ is given in \eqref{eq:PDFdistance_m_downlink}.

\bibliographystyle{IEEEtran}
\bibliography{IEEEabrv,Bibliography}

\end{document}